\documentclass{appolb}
\usepackage{graphicx}
\usepackage{amsmath,amssymb,amsfonts}
\usepackage{array}
\usepackage{url}
\usepackage{hyperref}
\usepackage{subfig}
\usepackage{multirow}
\usepackage{float}
\usepackage{lineno}
\usepackage{booktabs}% for better rules in the table
\usepackage{xspace}
\usepackage[usenames,dvipsnames]{color}
\definecolor{kugray5}{RGB}{224,224,224}
\usepackage{adjustbox}
\usepackage{booktabs}
\usepackage{multirow}
\usepackage{rotating,tabularx}
\usepackage{lscape}
\newcommand{\Kslxi}{$K^{0}_{s}$, $\Lambda$, and $\Xi^{-}$}
\newcommand{\Sqsn}{\mbox{$\sqrt{s_{NN}}$}\xspace}
\newcommand{\bef}{\begin{figure}}
\newcommand{\eef}{\end{figure}}
\newcommand{\bc}{\begin{center}}
\newcommand{\ec}{\end{center}}

\newcommand{\be}{\begin{equation}}
\newcommand{\ee}{\end{equation}}
\newcommand{\bea}{\begin{eqnarray}}
\newcommand{\eea}{\end{eqnarray}}
\begin{document}
\title{Transverse momentum $p_{\mathrm{T}}$ spectra of strange particles production in different collisions at $\Sqsn= 2.76, 5.02,$ and $7$ TeV} %at \textit{CMS}
\author{Hayam Yassin \thanks{Corresponding Author: hiam$\_$hussien@women.asu.edu.eg}, Eman R. Abo Elyazeed\thanks{eman.reda@women.asu.edu.eg}
 \address{Physics Department, Faculty of Women for Arts, Science and Education, Ain Shams University, 11577 Cairo, Egypt}
}
\maketitle
\begin{abstract}
We analyse the transverse momentum $p_{\mathrm{T}}$ spectra of strange particles \Kslxi produced in $\textsf{Pb+Pb}$ collision at $\Sqsn= 2.76$ TeV, $\textsf{p+Pb}$ collision at $\Sqsn= 5.02$ TeV, and $\textsf{p+p}$ collision at $\Sqsn= 7$ TeV in different multiplicity events measured by the \textit{CMS} experiment at the Large Hadron Collider. The $p_{\mathrm{T}}$ spectra of strange particles are fitted by Tsallis statistics and Boltzmann statistics, respectively. The fitting parameters are studied as a function of the multiplicity events for all systems. The Tsallis temperature ($T_{\mathtt{Ts}}$), Boltzmann temperature ($T_{\mathtt{Boltz}}$), and radius of the system ($R$) increase with both the mass and strangeness number of the particle and also increase with the multiplicity events. The non-extensive parameter ($q$) decreases with the increase in the mass of the particle and also decrease with the increase in the multiplicity events which means that the system tends to thermodynamic stabilization. The extracted temperatures from the two statistics for the strange particles are exhibited a linear correlation.
\end{abstract}
%%%%%%%%%%%%%%%%%
\section{Introduction}
\label{intro}
Quark-gluon plasma (QGP) was created at heavy-ion collisions at the Relativistic Heavy-Ion Collider (RHIC) and the Large Hadron Collider (LHC) \cite{Bjorken:1982qr,Ullrich:2013qwa,Gyulassy:2004zy}. The heavy-ion experiments seek to study strongly interacting matter under extreme conditions of high density and /or high temperatures \cite{Khuntia:2017ite}. QGP was described as the deconfinement of the colliding hadrons which rapidly expands and cools down \cite{Ahmad:2013una}. "Hadronization" or phase transition from QGP to hadrons found at the combination of the quarks and gluons at special temperatures defined as critical temperatures. After cooling again, chemical freeze-out occur where the produced particles are stable \cite{Tawfik:2014eba}. One of the signatures for the creation of QGP is the enhancement of the strangeness number \cite{Rafelski:1994gi,ALICE:2017jyt}. Strange quarks are not produced into the reaction by the colliding nuclei. Therefore, any strange quarks or anti-quarks seen in experiments have been newly created from the kinetic energy of the colliding nuclei \cite{ALICE:2017jyt,Rafelski:1982pu}. Also, another important indication for the formation of QGP is the transverse momentum $p_{\mathrm{T}}$ spectra of the charged and strange particles \cite{Song:2010mg}. Because the $p_{\mathrm{T}}$ spectra can give information about the chemical freezeout - chemical potential and temperature - by utilizing many statistical models \cite{Adams:2003xp,Becattini:2004td,Mueller:1993rr,Mueller:1994jq,Mueller:1994gb,Schnedermann:1993ws,Tawfik:2017bul}. These statistical models describe the experimental measurements over a wide range of center-of-mass energies depending on different statistics.

Tsallis statistics \cite{Tsallis:1987eu,Biro:2008hz,22,23,24,25} success in fitting the experimental data \cite{Adare:2011vy,Khachatryan:2011tm} of the transverse momentum $p_{\mathrm{T}}$ spectra in high energy physics. The $p_{\mathrm{T}}$ spectra of all identified particles which measured in $\textsf{p+p}$ collisions at RHIC and LHC energies was fitted excellently by Tsallis statistics in Ref. \cite{Khandai:2013gva,Adare:2010fe,Sett:2014csa}. The Tsallis statistics with the transverse flow effect is used in the analysis of the $p_{\mathrm{T}}$ spectra of charged and strange particles in Ref. \cite{Khandai:2013fwa,Saraswat:2017gqt,21,24,25,Sett:2015lja}. Tsallis and Boltzmann statistics describe the $p_{\mathrm{T}}$ spectra of all identified particles measured in $\textsf{Au+Au}$ collision and $\textsf{Pb+Pb}$ collision at RHIC and LHC energies in Ref. \cite{Gao:2015qsq}. Also, the obtained temperatures from these previous fitting gave some useful correlations with each other. The aim of this paper is to obtain the correlation between the Tsallis and Boltzmann temperatures from the $p_{\mathrm{T}}$ spectra of strange particles at different collisions and energies.

The paper is organized as follows. In Sect. \ref{sec:2}, the Tsallis statistics is presented, which is used to describe the particle spectra. In Sect. \ref{sec:3}, we present the Boltzmann statistics which is used to describe the particle spectra. Then we discuss the results of the description of the $p_{\mathrm{T}}$ spectra of strange particles at different collisions and energies by using both statistics in Sect. \ref{sec:results}. Also, the dependence of the fitting parameters on the multiplicity events and particle mass is discussed in Sect. \ref{sec:results}. Finally, in Sect. \ref{sec:concl}  the conclusion of our results is presented.

%%%%%%%%%%%%%%%%%%%%%%%%%%%%%%%%%%%%%%%%%%%%%%%%%%%%%%%%%%%%%%%%%%%%%%%%%%%%%
%%%%%%%%%%%%%%%%%%%%%%%%%%%%%%%%%%%%%%%%%%%%%%%%%%%%%%%%

\section{Transverse momentum spectra}
\label{sec:1}
Transverse momentum $p_{\mathrm{T}}$ spectra of strange particles \Kslxi prod-uced in $\textsf{Pb+Pb}$ collision at $\Sqsn= 2.76$ TeV, $\textsf{p+Pb}$ collision at $\Sqsn= 5.02$ TeV, and $\textsf{p+p}$ collision at $\Sqsn= 7$ TeV which measured by the \textit{CMS} experiment will be discussed using Tsallis and Boltzmann statistics in the following Sects. \ref{sec:2} and \ref{sec:3}.

%%%%%%%%%%%%%%%%%%%%%%%%%%%%%%%%
\subsection{Tsallis statistics }
\label{sec:2}
The experimental measurements of the $p_{\mathrm{T}}$ spectra in high-energy collisions can be described by different formula of Tsallis statistics \cite{Tsallis:1987eu,Tsallis:1998ws,Cleymans:2012ya,46}. The total number of particles is given by
\begin{eqnarray}
\label{eq4}
N = \frac{g V}{(2\pi)^3} \int_0^\infty \left[1+(q-1)\left(\frac{E-\mu}{T_\mathrm{Ts}}\right) \right]^{\frac{q}{1-q}} d{p^3},
\end{eqnarray}
 where,  $T_\mathrm{Ts}$ is the Tsallis temperature, $q$ is the non-extensive parameter, $E$  is the energy, $p$ is the pressure, $g$ is the degeneracy factor, $V$ is the system volume, and  $\mu$ is the chemical potential. The momentum distribution \cite{Cleymans:2015lxa} can be obtained as
\begin{eqnarray}
\label{eq5}
E \frac{d^3N}{dp^3} = \frac{g V E}{(2\pi)^3} \left[1+(q-1)\left(\frac{E-\mu}{T_\mathrm{Ts}}\right)\right]^{\frac{q}{1-q}}.
\end{eqnarray}
In terms of the rapidity ($y$) and the transverse mass ($\;m_{\mathrm{T}}=\sqrt{\;p_{\mathrm{T}}^2 + m^2}$), energy can be written as $E=m_{\mathrm{T}} \cosh y$, so at mid-rapidity $ y=0 $ and $\mu \approx 0$, Eq. \ref{eq5} \cite{47} becomes

\begin{eqnarray}
\label{eq6}
\left.\frac{1}{2\pi p_{\mathrm{T}}}\frac{d^2N}{dp_{\mathrm{T}}dy}\right|_{y=0}=\frac{g V m_{\mathrm{T}}}{(2\pi)^3} \left[1+(q-1)\left(\frac{m_{\mathrm{T}}}{T_\mathrm{Ts}}\right)\right]^{\frac{q}{1-q}}.
\end{eqnarray}
%%%%%%%%%%%%%%%%%%%%%%%%%%%%%%%%

\subsection{Boltzmann statistics }
\label{sec:3}
According to Boltzmann statistics \cite{2}, the number of particles can be written as
\begin{eqnarray}
\label{eq1}
N = \frac{g V}{(2\pi)^3} \int_0^\infty \frac{d{p^3}}{\exp\left(\frac{E-\mu}{T_\mathrm{Boltz}}\right)} ,
\end{eqnarray}
where $T_\mathrm{Boltz}$ is the Boltzmann temperature. The momentum distribution \cite{2,3,Meng:2009zzc} can be obtained as
\begin{eqnarray}
\label{eq2}
E \frac{d^3N}{dp^3} = \frac{g V E}{(2\pi)^3} \exp\left(\frac{\mu-E}{T_\mathrm{Boltz}}\right).
\end{eqnarray}

The momentum distribution at mid-rapidity $ y=0 $ and $\mu \approx 0$ can be given by

\begin{eqnarray}
\label{eq3}
\left.\frac{1}{2\pi p_{\mathrm{T}}}\frac{d^2N}{dp_{\mathrm{T}}dy}\right|_{y=0}=\frac{g V m_{\mathrm{T}}}{(2\pi)^3} \exp\left(-\frac{m_{\mathrm{T}}}{T_\mathrm{Boltz}}\right).
\end{eqnarray}

 %%%%%%%%%%%%%%%%%%%%%%%%%%%%%%%%%%%%%%%%%%%%%%%%

\begin{figure}[h]
\begin{center}
\includegraphics[scale=0.65]{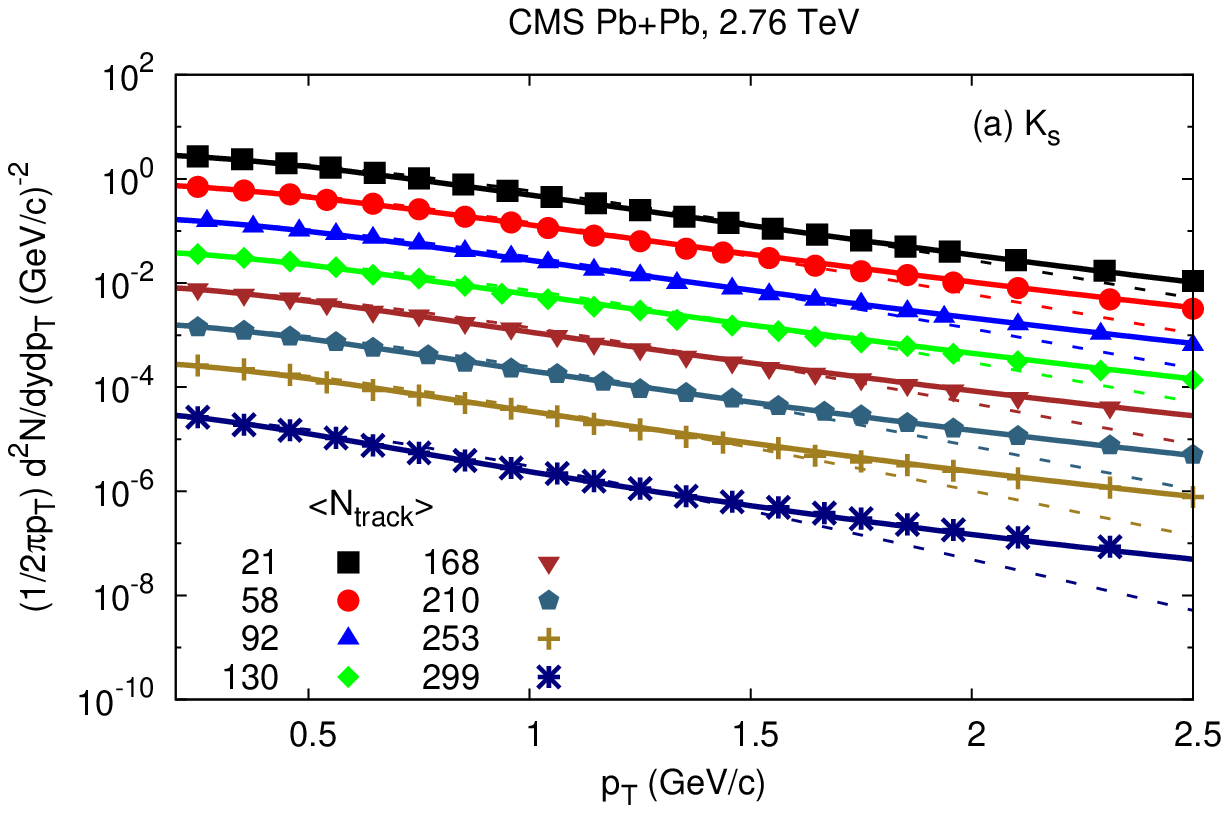}
\includegraphics[scale=0.65]{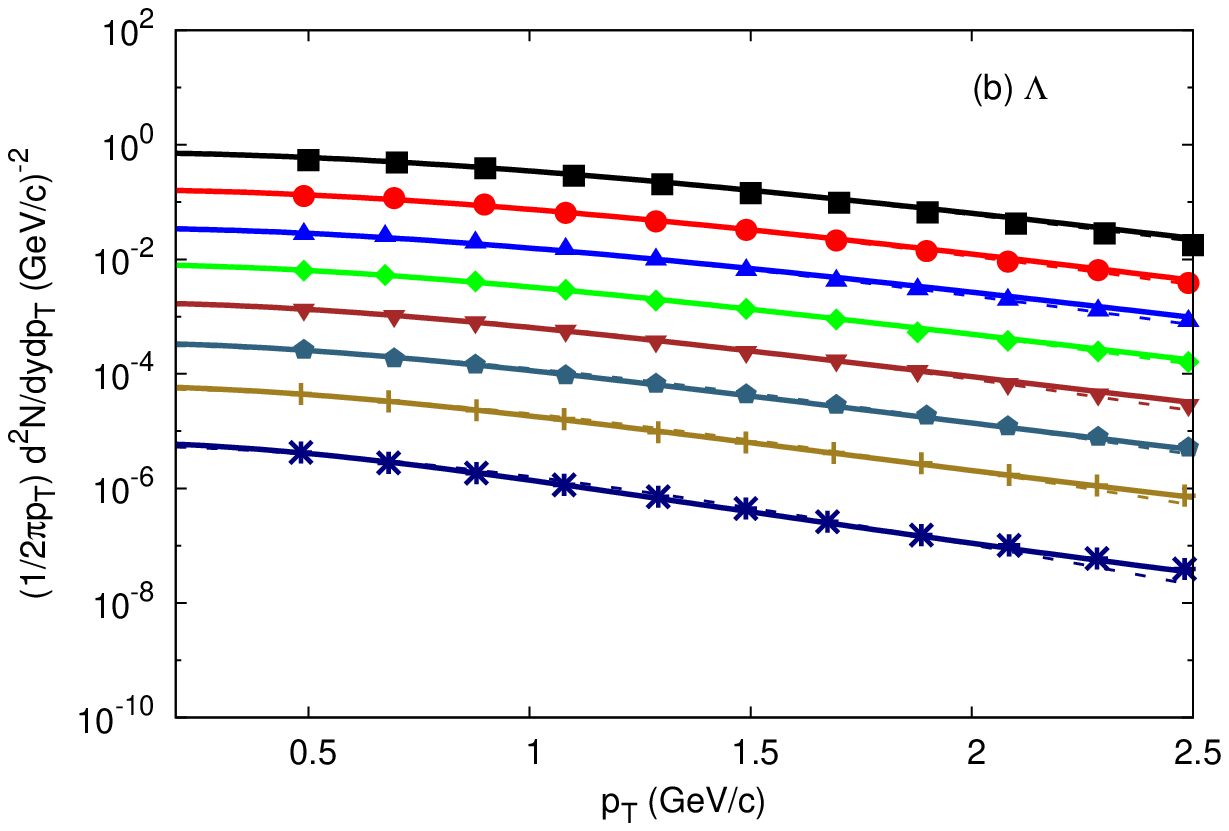}
\includegraphics[scale=0.65]{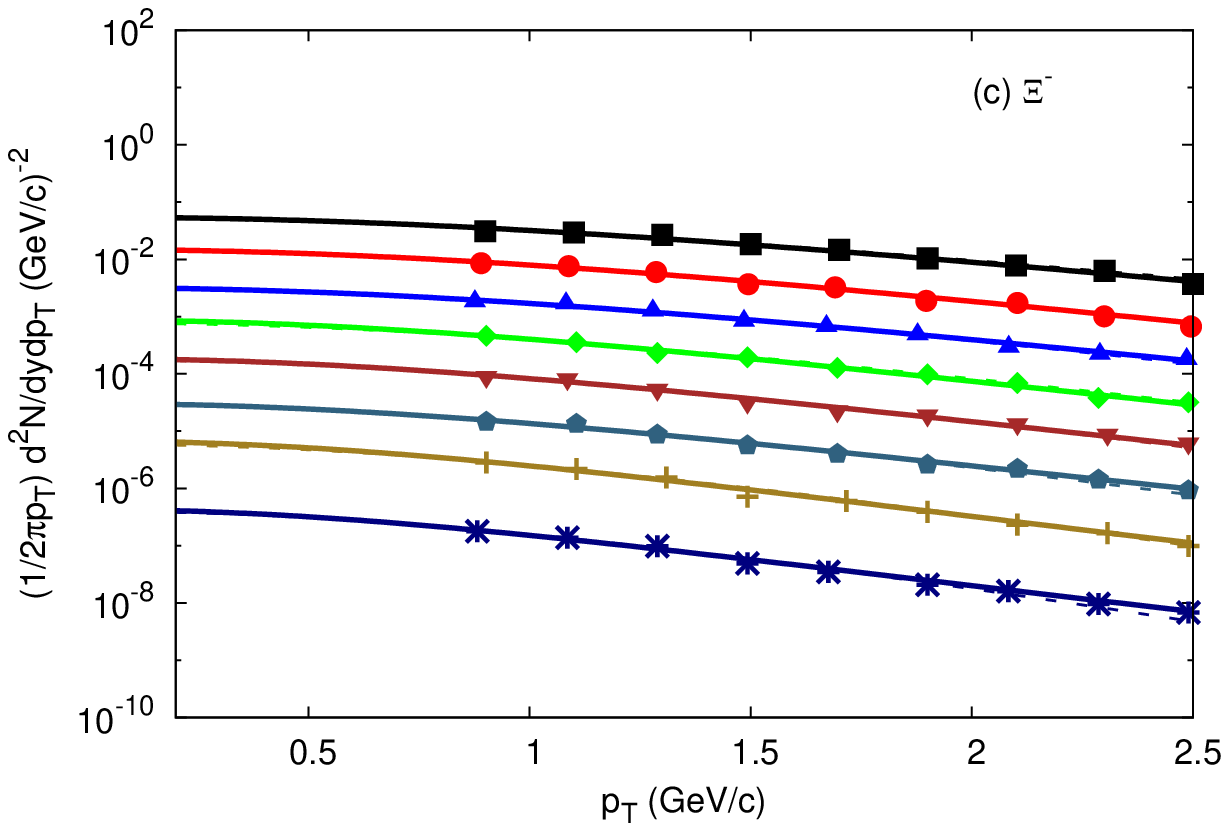}
\caption{(Color online) Transverse momentum distributions of the strange particles \Kslxi for $\textsf{Pb+Pb}$ collision at $\Sqsn= 2.76$ TeV measured in \textit{CMS} \cite{Khachatryan:2016yru} experiment (symbols) are compared with calculations from Tsallis statistics (solid curves) using Eq.~\ref{eq6} and with Boltzmann statistics (dashed curves) using Eq.~\ref{eq3} for different multiplicity intervals.}
\label{fit:Both:2.76}
\end{center}
\end{figure}
%%%%%%%%%%%%%%%%%%%%%%%%%%%%%%%%
 %%%%%%%%%%%%%%%%%%%%%%%%%%%%%%%%
\begin{figure}[h]
\begin{center}
\includegraphics[scale=0.65]{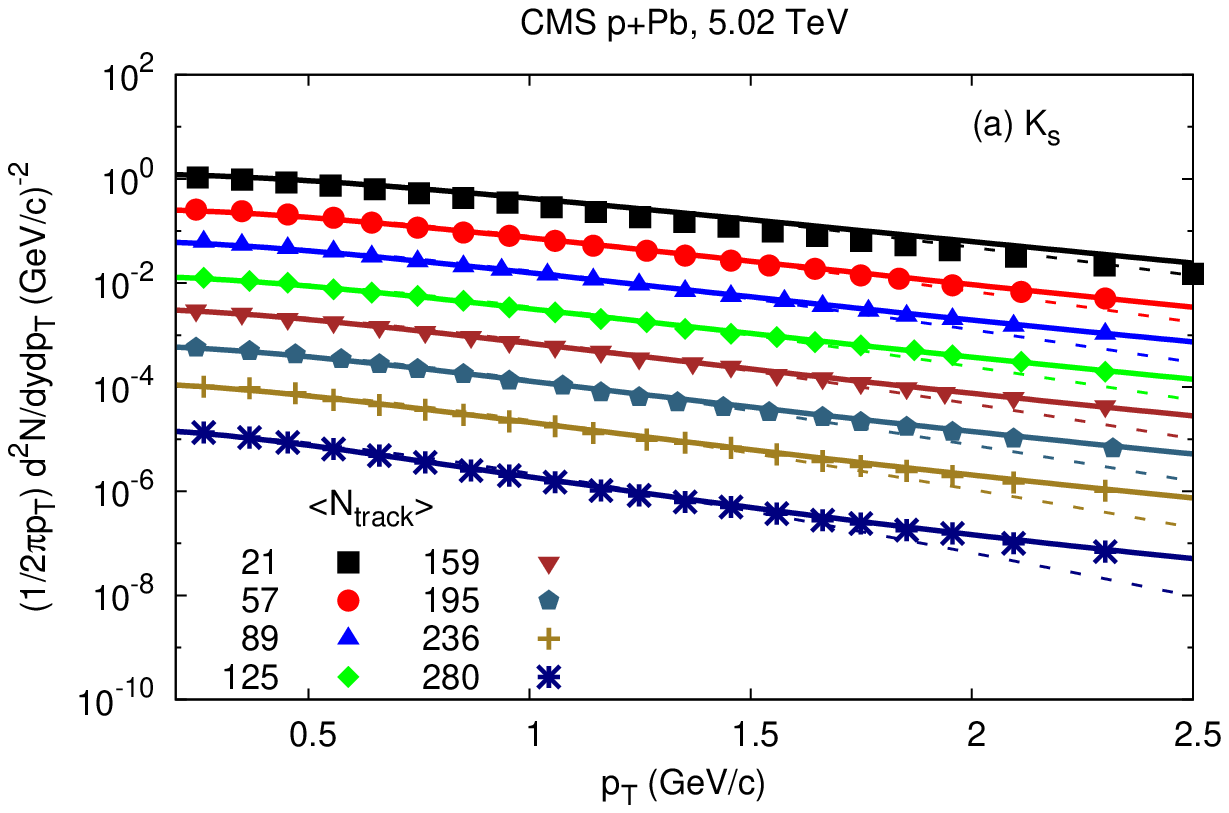}
\includegraphics[scale=0.65]{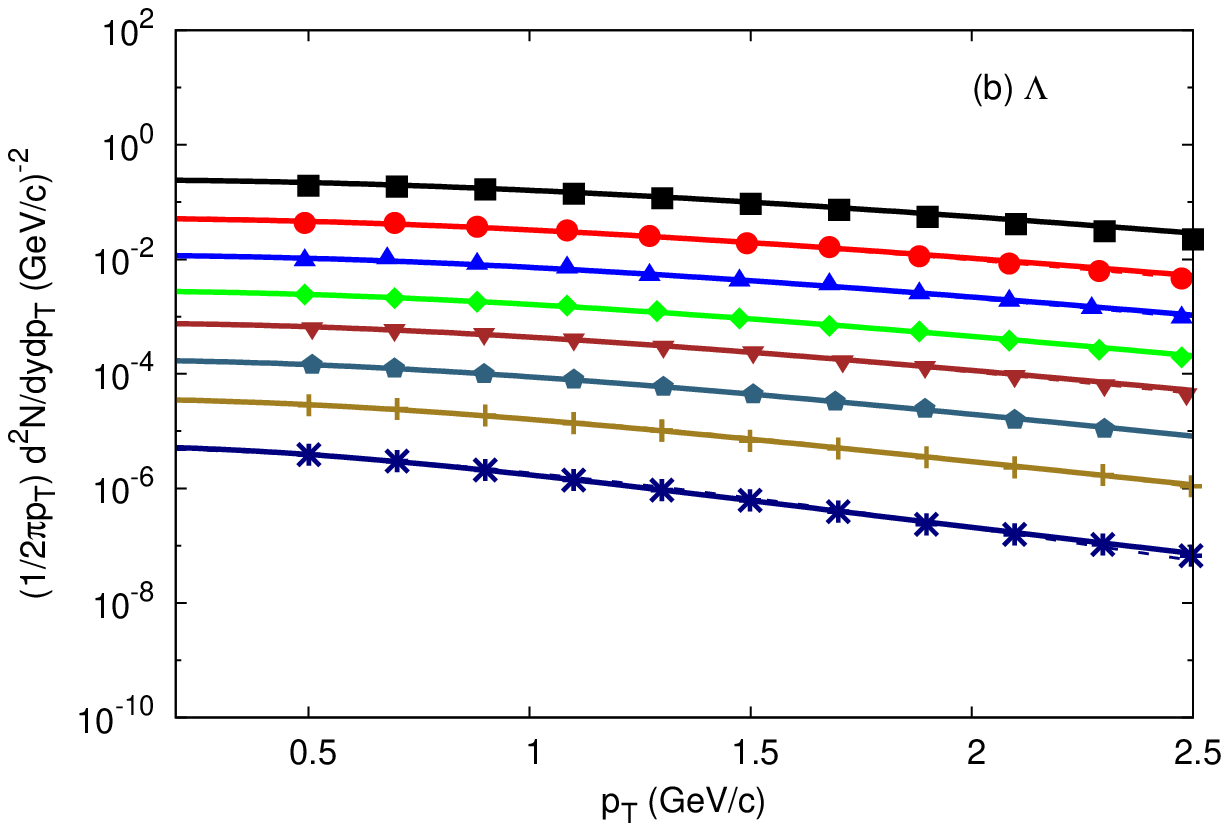}
\includegraphics[scale=0.65]{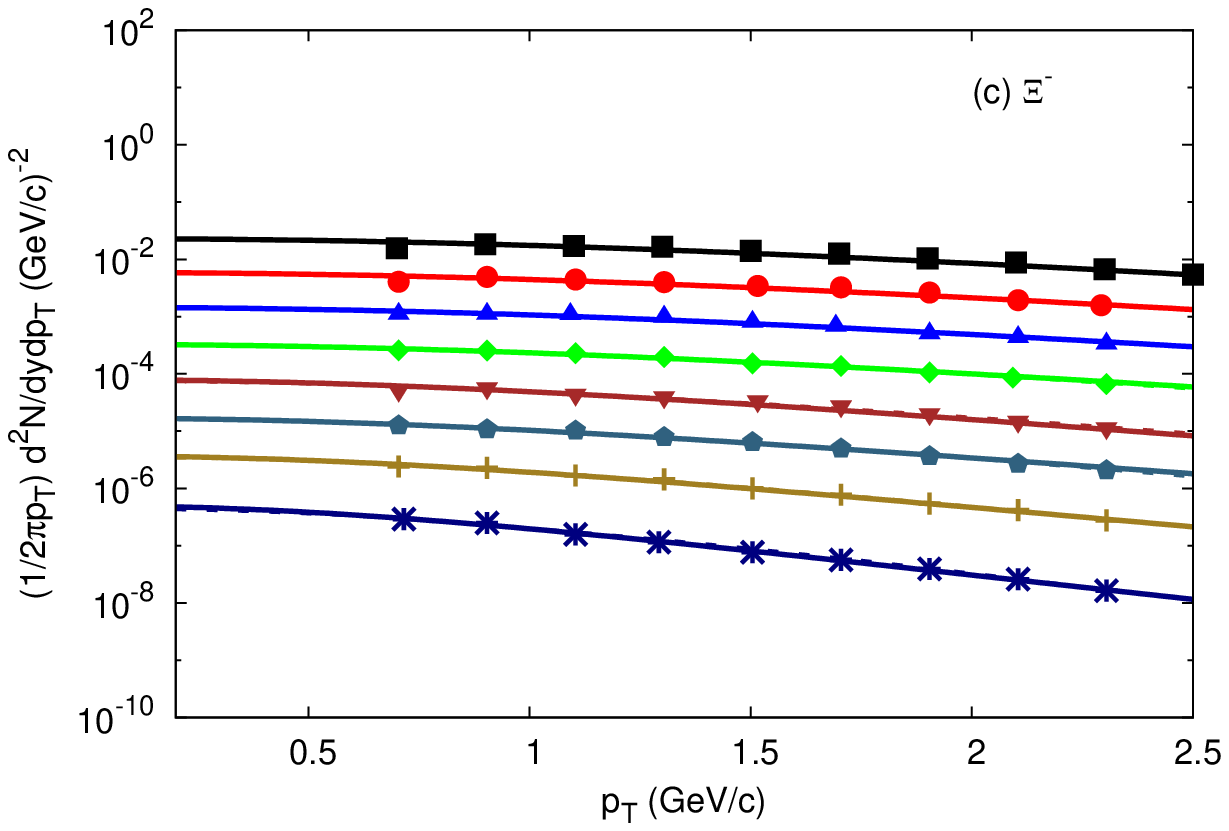}
\caption{(Color online) Transverse momentum distributions of the strange particles \Kslxi for $\textsf{p+Pb}$ collision at $\Sqsn= 5.02$ TeV measured in \textit{CMS} \cite{Khachatryan:2016yru} experiment (symbols) are compared with calculations from Tsallis statistics (solid curves)  using Eq.~\ref{eq6} and with Boltzmann statistics (dashed curves) using Eq.~\ref{eq3} for different multiplicity intervals.}
\label{fit:Both:5.02}
\end{center}
\end{figure}
%%%%%%%%%%%%%%%%%%%%%%%%%%%%%

%%%%%%%%%%%%%%%%%%%%%%%%%%%%%%%%
\begin{figure}[t!]
\begin{center}
\includegraphics[scale=0.65]{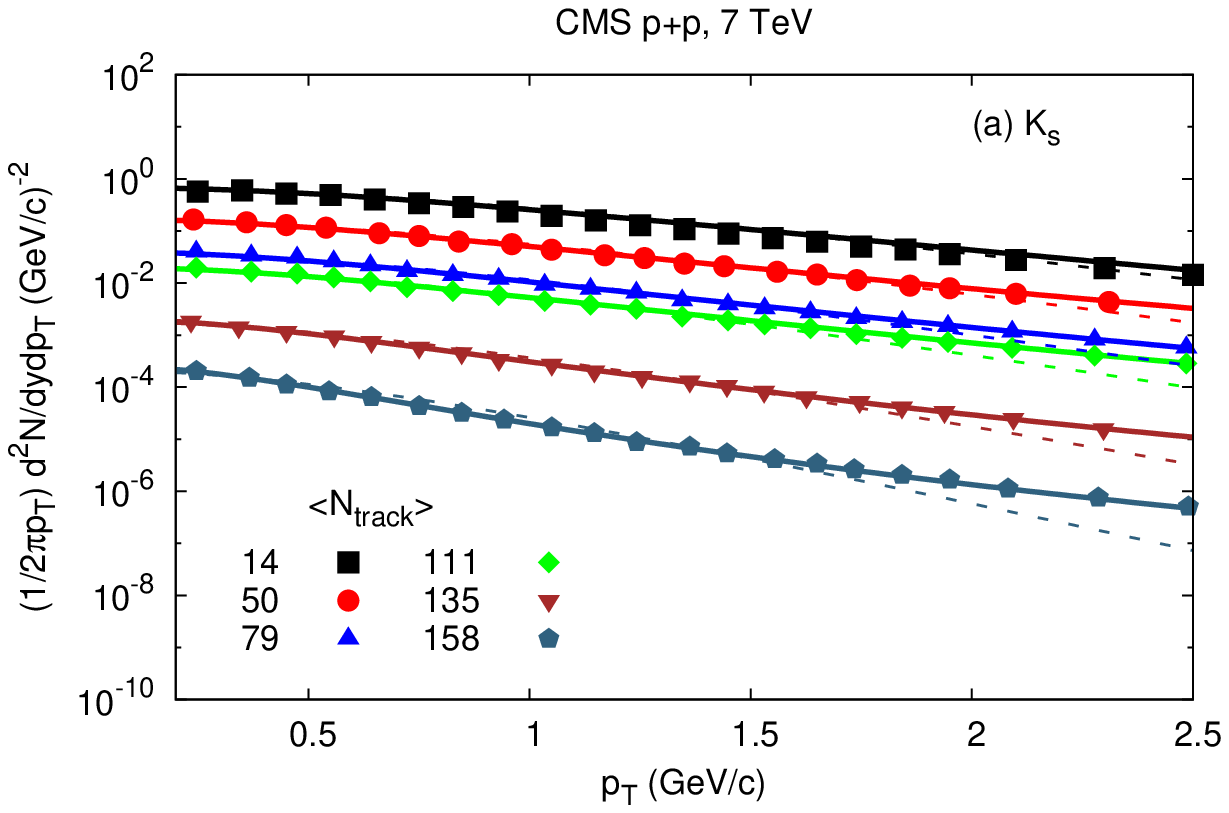}
\includegraphics[scale=0.65]{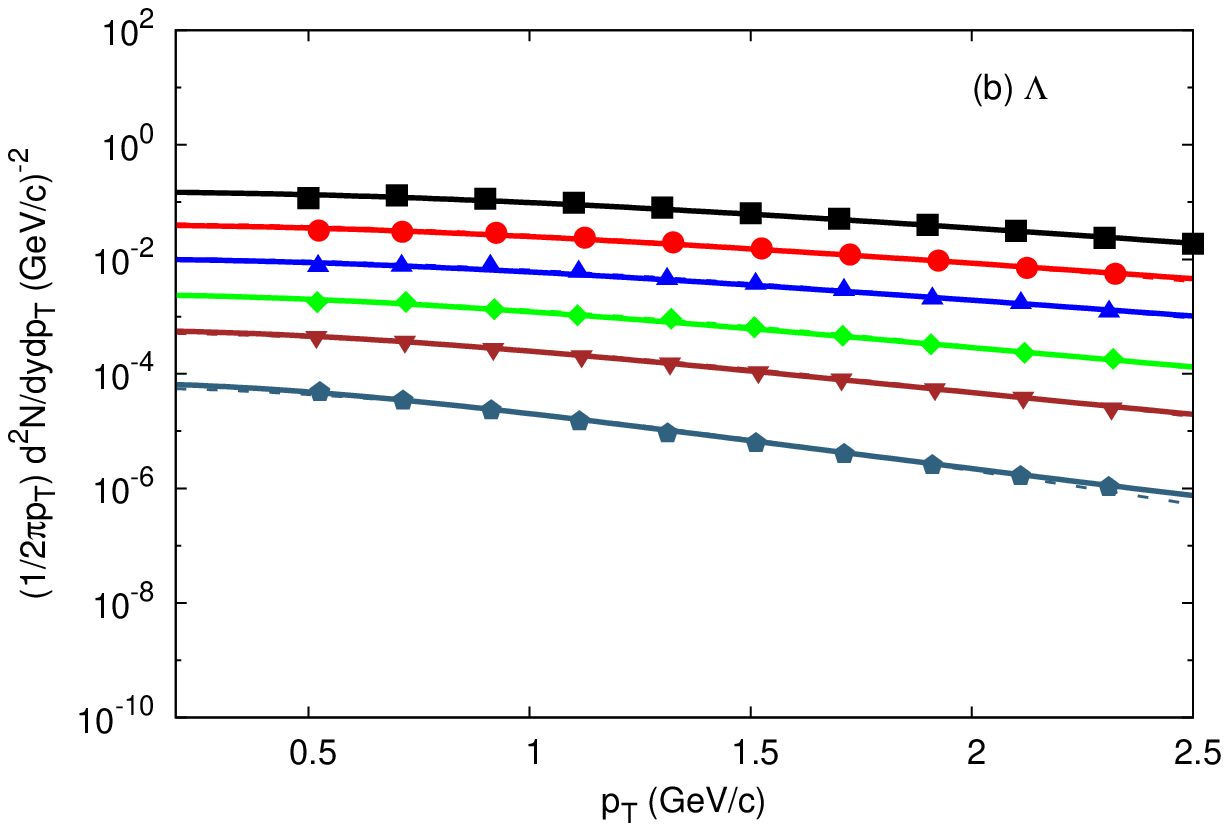}
\includegraphics[scale=0.65]{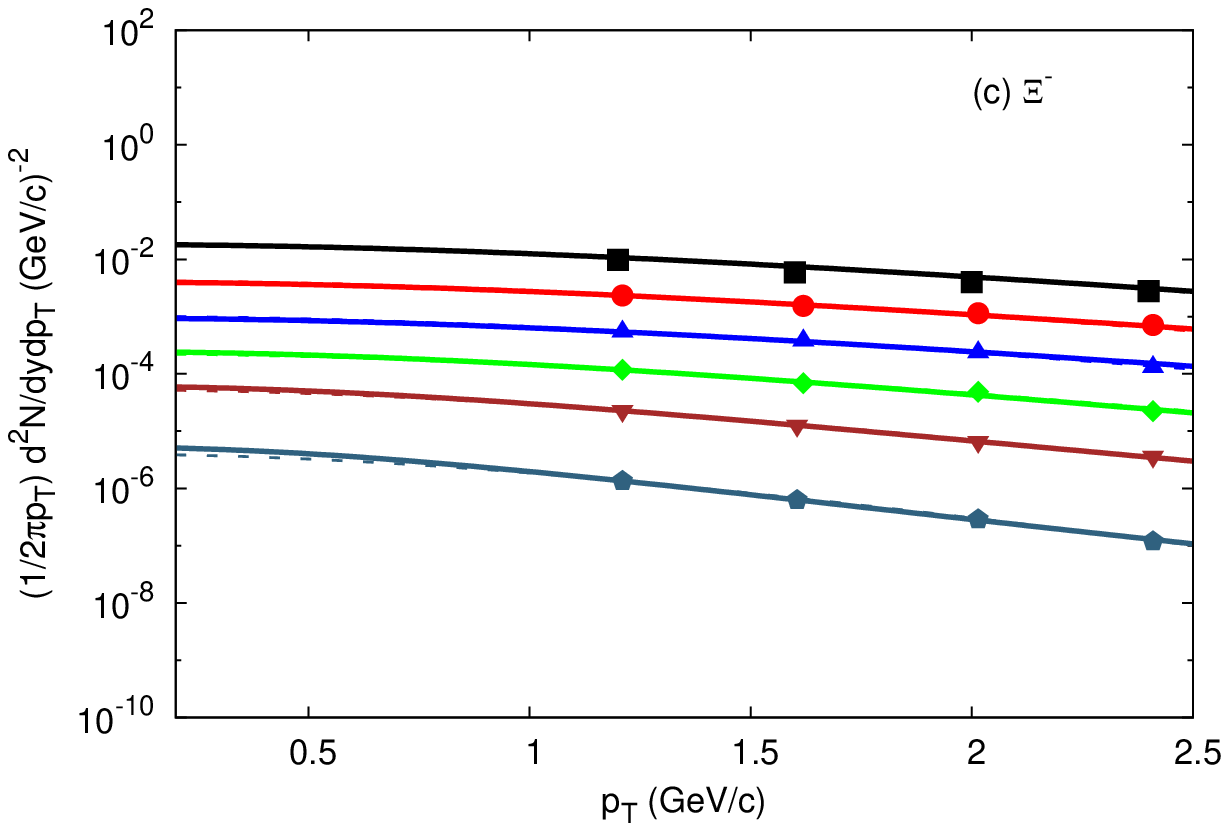}
\caption{(Color online) Transverse momentum distributions of the strange particles \Kslxi for $\textsf{p+p}$ collision at $\Sqsn= 7$ TeV measured in \textit{CMS} \cite{Khachatryan:2016yru} experiment (symbols) are compared with calculations from Tsallis statistics (solid curves)  using Eq.~\ref{eq6} and with Boltzmann statistics (dashed curves) using Eq.~\ref{eq3} for different multiplicity intervals.}
\label{fit:Both:7}
\end{center}
\end{figure}
%%%%%%%%%%%%%%%%%%%%%%%%%%%%%%%%

 %%%%%%%%%%%%%%%%%%%%%%%%%%%%%%%%%%%%%%%%%%%%%%%%
\section{Results and Discussion}
\label{sec:results}
 The transverse momentum distributions of the strange particles \Kslxi in $\textsf{Pb+Pb}$ collision at $\Sqsn= 2.76$ TeV, $\textsf{p+Pb}$ collision at $\Sqsn= 5.02$ TeV, and $\textsf{p+p}$ collision at $\Sqsn= 7$ TeV \cite{Khachatryan:2016yru} in different multiplicity intervals are fitted using two different kinds of statistics. Boltzmann statistics using Eq.~\ref{eq3} and Tsallis statistics using Eq.~\ref{eq6}. These are shown in Figs. \ref{fit:Both:2.76}, \ref{fit:Both:5.02}, and \ref{fit:Both:7}. The fitting parameters are listed in Tables \ref{Tab1}, \ref{Tab2}, and \ref{Tab3}.

 Figure~\ref{fit:Both:2.76} shows the $p_{\mathrm{T}}$ spectra of the strange particles, (a) $K_s^0$, (b) $\Lambda$, and (c) $\Xi^-$, produced in $\textsf{Pb+Pb}$ collision at $\Sqsn= 2.76$ TeV with different multiplicity intervals. The experimental data of the \textit{CMS} experiment \cite{Khachatryan:2016yru} are represented by symbols. The experimental data are divided into classes based on the multiplicity intervals $N_{\mathrm{trk}}^{\mathrm{offline}}$ in the mid-rapidity range $\vert y \vert<1.0$. The corresponding averaged multiplicity $<N_{\mathrm{track}}> = 21, 58, 92, 130, 168, 210, 253$ and $299$ \cite{Chatrchyan:2013nka}. The solid and dashed curves are the calculated results using the Tsallis statistics (Eq.~\ref{eq6}) and the Boltzmann statistics (Eq.~\ref{eq3}), respectively. Here, we concentrate on the smallest $p_{\mathrm{T}}$ region. The resulting fit parameters are given in Table \ref{Tab1}. We notice that our two results for $\Lambda$ and $\Xi^-$ particles are in good agreement with the experimental data of $\textsf{Pb+Pb}$ collisions at $\Sqsn= 2.76$ TeV for all multiplicity intervals. For $K_s^0$ particle, also there is a good agreement only with the results obtained from Tsallis statistics but with the results obtained from Boltzmann statistics, the good agreement decreases with the increase in the multiplicity classes.

%%%%%%%%%%%%%%%%%%%%%%%%%%%%%%%%%%%
\begin{landscape}
\begin{table*}[t]
\small
\begin{center}
\caption{Parameters deduced for the fitting of Tsallis and Boltzmann transverse momentum distributions at different multiplicity intervals for $\textsf{Pb+Pb}$ collision at $\sqrt{s_{\mathtt{NN}}}=2.76$ TeV, Fig. \ref{fit:Both:2.76}.}
\label{Tab1}
\begin{tabular}{ccccccccc}
%\toprule
\hline\noalign{\smallskip}
Particle & $<N_{\mathrm{track}}>$ & $T_{\mathtt{Ts}}$ [MeV] & $q$ & $R$ [fm] & $\chi^2/\mathrm{dof}$ & $T_{\mathtt{Boltz}}$ [MeV] & $R$ [fm] & $\chi^2/\mathrm{dof}$\\
\noalign{\smallskip}\hline\noalign{\smallskip}
  $K_s^0$  & $21$ & $87.169 \pm 0.105$ & $1.15 \pm 0.001$ & $0.154 \pm 0.026$ & $0.023$ & $198.808 \pm 1.672$ & $0.07 \pm 0.021$ & $0.13$\\
    		 & $58$ & $133.113 \pm 0.378$ & $1.1224 \pm 0.003$ & $0.211 \pm 0.044$ & $0.047$ & $220.873 \pm 1.47$ & $0.136 \pm 0.036$ & $0.1$\\
   		 & $92$ & $139.857 \pm 0.131$ & $1.1189 \pm 0.001$ & $0.358 \pm 0.052$ & $0.015$ & $228.78 \pm 1.46$ & $0.236 \pm 0.062$ & $0.09$\\
  		 & $130$ & $147.611 \pm 0.283$ & $1.1158 \pm 0.002$ & $0.589 \pm 0.107$ & $0.03$ & $238.526 \pm 1.749$ & $0.394 \pm 0.107$ & $0.1$ \\
 		 & $168$ & $165.663 \pm 0.418$ & $1.1017 \pm 0.003$ & $0.894 \pm 0.175$ & $0.038$ & $248.365 \pm 1.7$ & $0.642 \pm 0.169$ & $0.091$ \\
	     & $210$ & $170.876 \pm 0.345$ & $1.1007 \pm 0.003$ & $1.421 \pm 0.257$ & $0.03$ & $248.671 \pm 1.624$ & $1.051 \pm 0.273$ & $0.087$  \\
 	     & $253$ & $181.643 \pm 0.294$ & $1.0941 \pm 0.002$ & $2.226 \pm 0.371$ & $0.023$ & $249.084 \pm 1.053$ & $1.741 \pm 0.391$ & $0.056$ \\
	     & $299$ & $188.54 \pm 0.499$ & $1.079 \pm 0.004$ & $3.381 \pm 0.622$ & $0.031$ & $256.769 \pm 2.215$ & $2.65 \pm 0.737$ & $0.108$ \\
\noalign{\smallskip}\hline\noalign{\smallskip}
  $\Lambda$ & $21$ & $91.274 \pm 0.084$ & $1.11 \pm 0.0003$ & $0.193 \pm 0.034$ & $0.056$ & $249.745 \pm 1.111$ & $0.046 \pm 0.013$ & $0.233$ \\
    		 & $58$ & $160.973 \pm 0.093$ & $1.09 \pm 0.0003$ & $0.166 \pm 0.024$ & $0.03$ & $289.72 \pm 0.944$ & $0.082 \pm 0.02$ & $0.15$ \\
   		 & $92$ & $187.226 \pm 0.27$ & $1.08 \pm 0.001$ & $0.242 \pm 0.046$ & $0.07$ & $304.703 \pm 1.244$ & $0.137 \pm 0.036$ & $0.181$ \\
  		 & $130$ & $229.993 \pm 0.164$ & $1.064 \pm 0.001$ & $0.321 \pm 0.047$ & $0.032$ & $309.954 \pm 0.359$ & $0.235 \pm 0.04$ & $0.05$\\
 		 & $168$ & $274.28 \pm 0.412$ & $1.0429 \pm 0.001$ & $0.441 \pm 0.082$ & $0.064$ & $329.929 \pm 0.511$ & $0.366 \pm 0.068$ & $0.064$\\
	     & $210$ & $322.756 \pm 1.085$ & $1.0245 \pm 0.003$ & $0.605 \pm 0.143$ & $0.132$ & $335.32 \pm 1.454$ & $0.597 \pm 0.156$ & $0.178$ \\
 	     & $253$ & $343.331 \pm 1.146$ & $1.0086 \pm 0.003$ & $0.957 \pm 0.179$ & $0.13$ & $340.261 \pm 1.479$ & $0.982 \pm 0.255$ & $0.176$\\
	     & $299$ & $362.632 \pm 0.633$ & $1.0067 \pm 0.002$ & $1.488 \pm 0.279$ & $0.066$ & $360.228 \pm 0.887$ & $1.515 \pm 0.321$ & $0.095$\\
\noalign{\smallskip}\hline\noalign{\smallskip}
  $\Xi^-$  & $21$ & $156.108 \pm 0.513$ & $1.09 \pm 0.001$ & $0.042 \pm 0.011$ & $0.246$ & $290.041 \pm 1.119$ & $0.019 \pm 0.165$ & $0.284$ \\
    		 & $58$ & $188.975 \pm 0.719$ & $1.07 \pm 0.001$ & $0.081 \pm 0.023$ & $0.282$ & $309.878 \pm 1.633$ & $0.042 \pm 0.12$ & $0.367$\\
   		 & $92$ & $257.487 \pm 1.258$ & $1.0599 \pm 0.003$ & $0.085 \pm 0.025$ & $0.312$ & $340.112 \pm 2.124$ & $0.063 \pm 0.075$ & $0.402$\\
  		 & $130$ & $265.875 \pm 1.243$ & $1.052 \pm 0.002$ & $0.151 \pm 0.043$ & $0.3$ & $349.961 \pm 2.211$ & $0.11 \pm 0.047$ & $0.394$ \\
 		 & $168$ & $286.238 \pm 0.784$ & $1.0441 \pm 0.002$ & $0.231 \pm 0.055$ & $0.172$ & $379.967 \pm 1.907$ & $0.164 \pm 0.035$ & $0.292$ \\
	     & $210$ & $365.428 \pm 1.569$ & $1.0248 \pm 0.003$ & $0.27 \pm 0.072$ & $0.238$ & $390.23 \pm 2.162$ & $0.257 \pm 0.02$ & $0.314$ \\
 	     & $253$ & $377.162 \pm 1.579$ & $1.0168 \pm 0.003$ & $0.437 \pm 0.116$ & $0.232$ & $400.073 \pm 2.165$ & $0.414 \pm 0.013$ & $0.301$\\
	     & $299$ & $425.295 \pm 2.719$ & $1.0152 \pm 0.005$ & $0.595 \pm 0.176$ & $0.322$ & $470.406 \pm 3.561$ & $0.533 \pm 0.005$ & $0.368$ \\
\noalign{\smallskip}\hline
%\bottomrule
\end{tabular}
\vspace{0mm}
\end{center}
\vspace{0cm}
\end{table*}
\end{landscape}
%%%%%%%%%%%%%%%%%%%%%%%%%%%%%%%%%%%
 %%%%%%%%%%%%%%%%%%%%%%%%%%%%%%%%%%%
\begin{landscape}
\begin{table*}[t]
\small
\begin{center}
\caption{Parameters deduced for the fitting of Tsallis and Boltzmann transverse momentum distributions at different multiplicity intervals for $\textsf{p+Pb}$ collision collision at $\sqrt{s_{\mathtt{NN}}}=5.02$ TeV, Fig. \ref{fit:Both:5.02}.}
\label{Tab2}
\begin{tabular}{ccccccccc}
%\toprule
\hline\noalign{\smallskip}
Particle & $<N_{\mathrm{track}}>$ & $T_{\mathtt{Ts}}$ [MeV] & $q$ & $R$ [fm] & $\chi^2/\mathrm{dof}$ & $T_{\mathtt{Boltz}}$ [MeV] & $R$[fm] & $\chi^2/\mathrm{dof}$\\
\noalign{\smallskip}\hline\noalign{\smallskip}
$K_s^0$ & $21$ & $133.898 \pm 0.295$ & $1.1307 \pm 0.002$ & $0.077 \pm 0.015$ & $0.035$ & $228.653 \pm 1.589$ & $0.049 \pm 0.013$ & $0.098$ \\
• & $57$ & $179.121 \pm 0.646$ & $1.1147 \pm 0.005$ & $0.118 \pm 0.026$ & $0.051$ & $256.887 \pm 1.525$ & $0.09 \pm 0.023$ & $0.078$ \\
• & $89$ & $199.756 \pm 0.687$ & $1.1083 \pm 0.005$ & $0.19 \pm 0.04$ & $0.046$ & $269.206 \pm 1.133$ & $0.153 \pm 0.034$ & $0.054$ \\
• & $125$ & $206.332 \pm 0.489$ & $1.1047 \pm 0.004$ & $0.32 \pm 0.059$ & $0.031$ & $278.893 \pm 1.433$ & $0.258 \pm 0.06$ & $0.064$ \\
• & $159$ & $218.697 \pm 0.833$  & $1.1021 \pm 0.006$ & $0.497 \pm 0.106$ & $0.048$ & $288.99 \pm 1.548$ & $0.409 \pm 0.096$ & $0.065$ \\
• & $195$ & $226.848 \pm 1.085$ & $1.1005 \pm 0.008$ & $0.81 \pm 0.185$ & $0.06$ & $294.12 \pm 1.769$ & $0.678 \pm 0.165$ & $0.072$ \\
• & $236$ & $255.94 \pm 1.87$ & $1.0787 \pm 0.013$ & $1.205 \pm 0.312$ & $0.087$ & $309.36 \pm 2.499$ & $1.063 \pm 0.282$ & $0.093$ \\
• & $280$ & $293.332 \pm 1.425$ & $1.0676 \pm 0.009$ & $1.869 \pm 0.419$ & $0.056$ & $334.762 \pm 1.591$ & $1.721 \pm 0.382$ & $0.055$ \\
\noalign{\smallskip}\hline\noalign{\smallskip}
  $\Lambda$ & $21$ & $172.072 \pm 0.124$ & $1.09 \pm 0.0004$ & $0.067 \pm 0.01$ & $0.036$ & $299.715 \pm 0.865$ & $0.035 \pm 0.008$ & $0.13$ \\
• & $57$ & $297.482 \pm 0.386$ & $1.0499 \pm 0.001$ & $0.065 \pm 0.011$ & $0.052$ & $369.708 \pm 0.797$ & $0.053 \pm 0.011$ & $0.083$ \\
• & $89$ & $373.144 \pm 0.253$ & $1.0249 \pm 0.001$ & $0.089 \pm 0.012$ & $0.024$ & $409.837 \pm 0.47$ & $0.082 \pm 0.013$ & $0.04$ \\
• & $125$ & $436.177 \pm 1.142$ & $1.0091 \pm 0.003$ & $0.127 \pm 0.026$ & $0.085$ & $430.532 \pm 1.625$ & $0.131 \pm 0.031$ & $0.127$ \\
• & $159$ & $453.51 \pm 0.919$ & $1.0065 \pm 0.003$ & $0.19 \pm 0.035$ & $0.064$ & $460.097 \pm 1.122$ & $0.188 \pm 0.037$ & $0.077$ \\
• & $195$ & $482.8 \pm 2.922$ & $1.0047 \pm 0.008$ & $0.292 \pm 0.077$ & $0.183$ & $470.728 \pm 3.661$ & $0.302 \pm 0.087$ & $0.243$ \\
• & $236$ & $501.749 \pm 2.302$ & $1.0034 \pm 0.007$ & $0.465 \pm 0.111$ & $0.135$ & $480.908 \pm 2.991$ & $0.487 \pm 0.13$ & $0.192$ \\
• & $280$ & $531.914 \pm 2.423$ & $1.0012 \pm 0.007$ & $0.747 \pm 0.176$ & $0.131$ & $535.578 \pm 2.905$ & $0.744 \pm 0.186$ & $0.157$ \\
\noalign{\smallskip}\hline\noalign{\smallskip}
  $\Xi^-$ & $21$ & $192.967 \pm 0.508$ & $1.083 \pm 0.001$ & $0.031 \pm 0.008$ & $0.176$ & $339.7 \pm 1.556$ & $0.016 \pm 0.009$ & $0.277$ \\
• & $57$ & $301.869 \pm 1.753$ & $1.0711 \pm 0.004$ & $0.034 \pm 0.01$ & $0.306$ & $420.171 \pm 1.799$ & $0.024 \pm 0.006$ & $0.217$ \\
• & $89$ & $396.942 \pm 2.066$ & $1.0684 \pm 0.005$ & $0.042 \pm 0.011$ & $0.232$ & $480.405 \pm 2.054$ & $0.036 \pm 0.019$ & $0.194$ \\
• & $125$ & $423.094 \pm 4.797$ & $1.049 \pm 0.013$ & $0.067 \pm 0.023$ & $0.547$ & $531.027 \pm 4.793$ & $0.054 \pm 0.017$ & $0.38$ \\
• & $159$ & $568.425 \pm 26.71$ & $1.0287 \pm 0.01$ & $0.084 \pm 0.022$ & $0.236$ & $580.505 \pm 3.308$ & $0.083 \pm 0.022$ & $0.222$ \\
• & $195$ & $610.815 \pm 6.597$ & $1.0219 \pm 0.02$ & $0.131 \pm 0.041$ & $0.382$ & $630.846 \pm 6.12$ & $0.128 \pm 0.039$ & $0.317$ \\
• & $236$ & $642.08 \pm 7.933$ & $1.0189 \pm 0.019$ & $0.202 \pm 0.066$ & $0.437$ & $662.482 \pm 9.601$ & $0.198 \pm 0.069$ & $0.051$ \\
• & $280$ & $660.022 \pm 10.78$ & $1.0139 \pm 0.04$ & $0.311 \pm 0.111$ & $0.558$ & $672.726 \pm 9.541$ & $0.308 \pm 0.104$ & $0.488$ \\
\noalign{\smallskip}\hline
%\bottomrule
\end{tabular}
\vspace{0mm}
\end{center}
\vspace{0mm}
\end{table*}
\end{landscape}
%%%%%%%%%%%%%%%%%%%%%%%%%%%%%%%%%%%

Figure~\ref{fit:Both:5.02} represents the $p_{\mathrm{T}}$ spectra of the strange particles, (a) $K_s^0$, (b) $\Lambda$, and (c) $\Xi^-$, produced in $\textsf{p+Pb}$ collisions at $\Sqsn= 5.02$ TeV with different multiplicity intervals. The experimental data of the \textit{CMS} experiment \cite{Khachatryan:2016yru} are represented by symbols. The experimental data are divided into classes based on the multiplicity intervals $N_{\mathrm{trk}}^{\mathrm{offline}}$ in the mid-rapidity range $\vert y \vert<1.0$. The corresponding averaged multiplicity $<N_{\mathrm{track}}> = 21, 57, 89, 125, 159, 195, 236$ and $280$ \cite{Chatrchyan:2013nka}. The solid and dashed curves are the calculated results using the Tsallis statistics (Eq.~\ref{eq6}) and the Boltzmann statistics (Eq.~\ref{eq3}), respectively. Here, we concentrate on the smallest $p_{\mathrm{T}}$ region. The resulting fit parameters are given in Table \ref{Tab2}. We notice that our two results for $\Lambda$ and $\Xi^-$ particles are in good agreement with the experimental data of $\textsf{p+Pb}$ collisions at $\Sqsn= 5.02$ TeV for all multiplicity intervals. For $K_s^0$ particle, also there is a good agreement only with the results obtained from Tsallis statistics but with the results obtained from Boltzmann statistics, the good agreement decrease with the increasing of the multiplicity classes.

Figure~\ref{fit:Both:7} depicts the $p_{\mathrm{T}}$ spectra of the strange particles, (a) $K_s^0$, (b) $\Lambda$, and (c) $\Xi^-$, produced in $\textsf{p+p}$ collision at $\Sqsn= 7$ TeV with different multiplicity intervals. The experimental data of the \textit{CMS} experiment \cite{Khachatryan:2016yru} are represented by symbols. The experimental data are divided into classes based on the multiplicity intervals $N_{\mathrm{trk}}^{\mathrm{offline}}$ in the mid-rapidity range $\vert y \vert<1.0$. The corresponding averaged multiplicity $<N_{\mathrm{track}}> = 14, 50, 79, 111, 135$ and $158$ \cite{Chatrchyan:2013nka}. The solid and dashed curves are the calculated results using the Tsallis statistics (Eq.~\ref{eq6}) and the Boltzmann statistics (Eq.~\ref{eq3}), respectively. Here, we concentrate on the smallest $p_{\mathrm{T}}$ region. The resulting fit parameters are given in Table \ref{Tab3}. We notice that our two results for $\Lambda$ and $\Xi^-$ particles are in good agreement with the experimental data of $\textsf{p+p}$ collisions at $\Sqsn= 7$ TeV for all multiplicity intervals. For $K_s^0$ particle, also there is a good agreement only with the results obtained from Tsallis statistics but with the results obtained from Boltzmann statistics, the good agreement decrease with the increase in the multiplicity intervals.

As seen from Tables \ref{Tab1}, \ref{Tab2}, and \ref{Tab3}, the value of $\chi^2/\mathrm{dof}$ is so small which represents the good quality of the fitting. Especially, the Tsallis statistics that gives excellent agreement with the experimental measurements for all multiplicity classes.
%As $\chi^2/\mathrm{dof}$ is below one, we don't need to modify Tsallis formula by adding the radial flow effect but we compare between results with those had radial flow effect \cite{Saraswat:2017gqt}.
Also, the fitting results by Tsallis statistics are better than those of Boltzmann statistics, especially at high range of $p_{\mathrm{T}}$.

 Furthermore, we have extracted the fitting parameters the non-extensive parameter $q$, the Tsallis temperature parameter $T_{\mathrm{Ts}}$, and the Boltzmann temperature parameter $T_{\mathrm{Boltz}}$ which inform us about the variations between Tsallis statistics and Boltzmann statistics. Also, there is another fitting parameter known as the radius $R$  as we assume the geometry of the fireball is spherical so $R=(3V/4\pi)^{1/3}$ which signifies the dimension of the system and is related to the normalization in the statistical distribution function which used in describing the particle yield or spectra \cite{50}.\\

\begin{figure*}[t!]
\begin{center}
\includegraphics[scale=0.65]{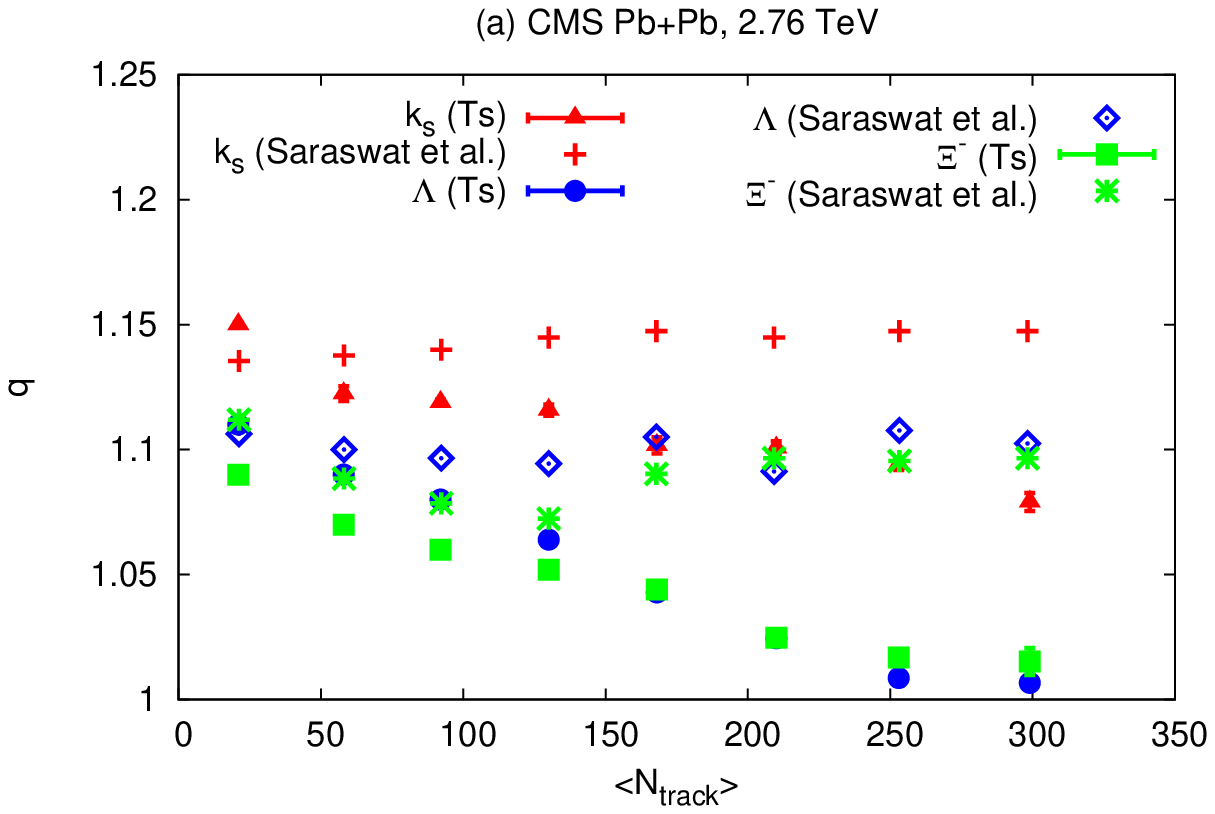}
\includegraphics[scale=0.65]{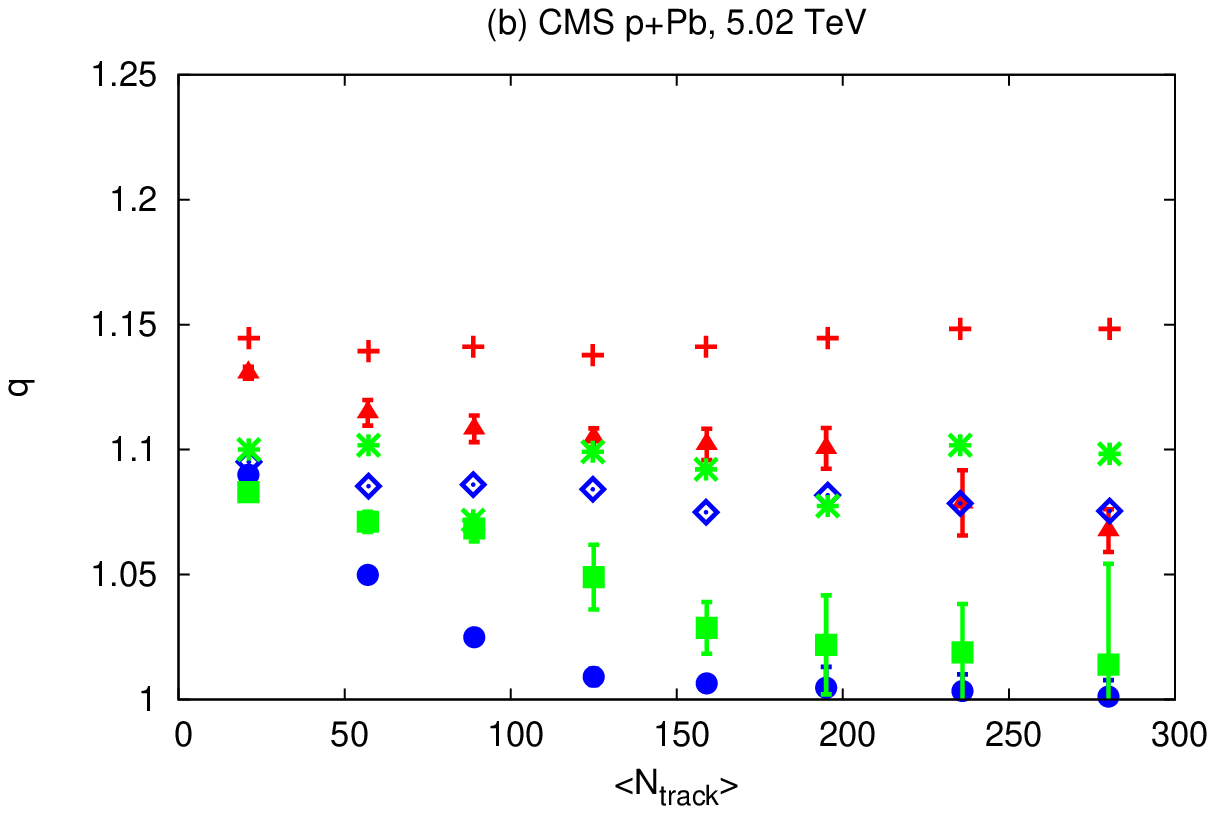}
\includegraphics[scale=0.65]{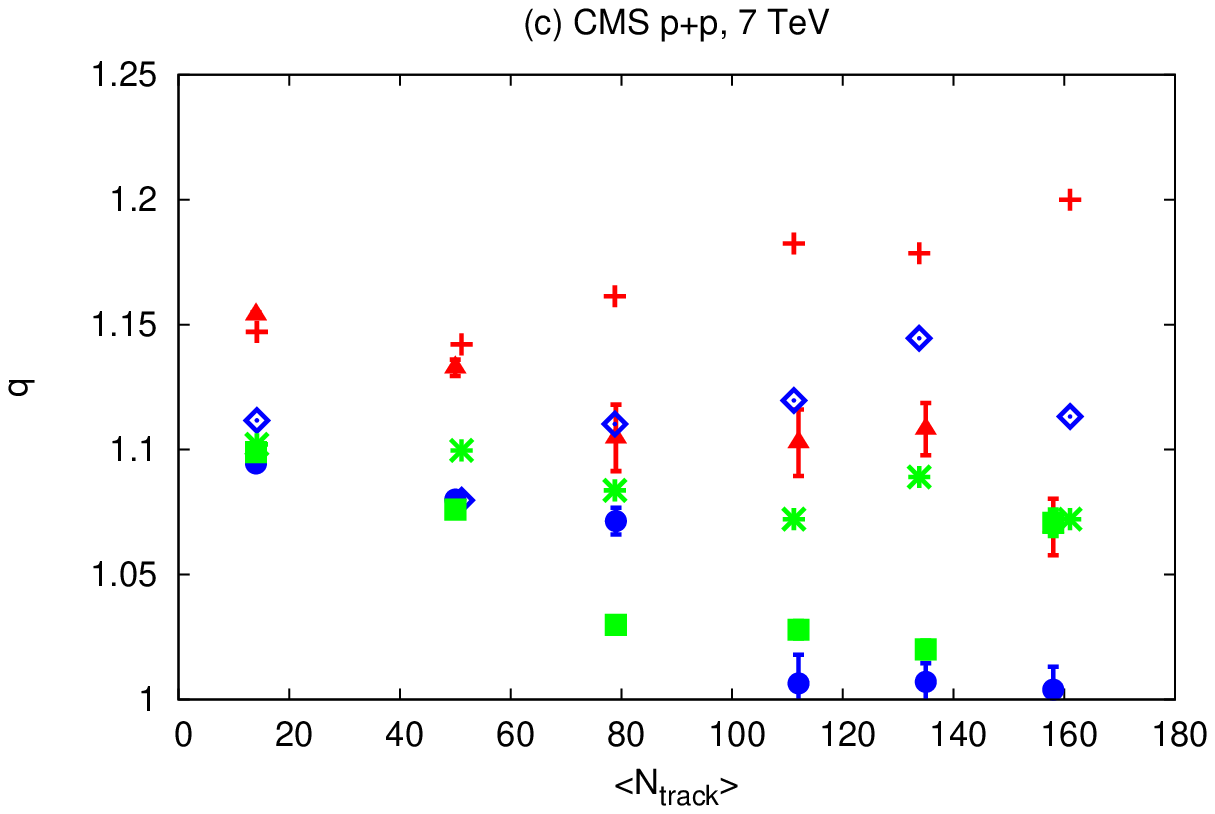}
\newline
\caption{(Color online) The fitting parameter $q$ as a function of the event multiplicity $<N_{\mathrm{track}}>$ for the strange particles \Kslxi which fitted using Tsallis and Boltzmann statistics for (a) $\textsf{Pb+Pb}$ collision at $\Sqsn=2.76$ TeV, (b) $\textsf{p+Pb}$ collision at $\Sqsn=5.02$ TeV, and (c) $\textsf{p+p}$ collision at $\Sqsn=7$ TeV.}
\label{comparision:q}
\end{center}
\end{figure*}

Figure~\ref{comparision:q} (a-c) depict the fitting parameter $q$ for the strange particles \Kslxi as a function of $<N_{\mathrm{track}}>$ produced in the $\textsf{Pb+Pb}, \textsf{p+Pb},$ and $\textsf{p+p}$ collisions at $\sqrt{s_{\mathtt{NN}}}=2.76, 5.02,$ and $7$ TeV, respectively. Results are obtained using Tsallis statistics and modified Tsallis statistics \cite{Saraswat:2017gqt}. The value of $q$ (our calculations) decreases with the increasing of both the particle mass and strangeness number and also with the increasing in the multiplicity events for all systems. Our results different from modified one in the effect of multiplicity on the non-extensivity parameter. This decreasing indicate that the system tend to be in equilibrium (thermodynamically).\\

\begin{figure*}[t!]
\begin{center}
\includegraphics[scale=0.65]{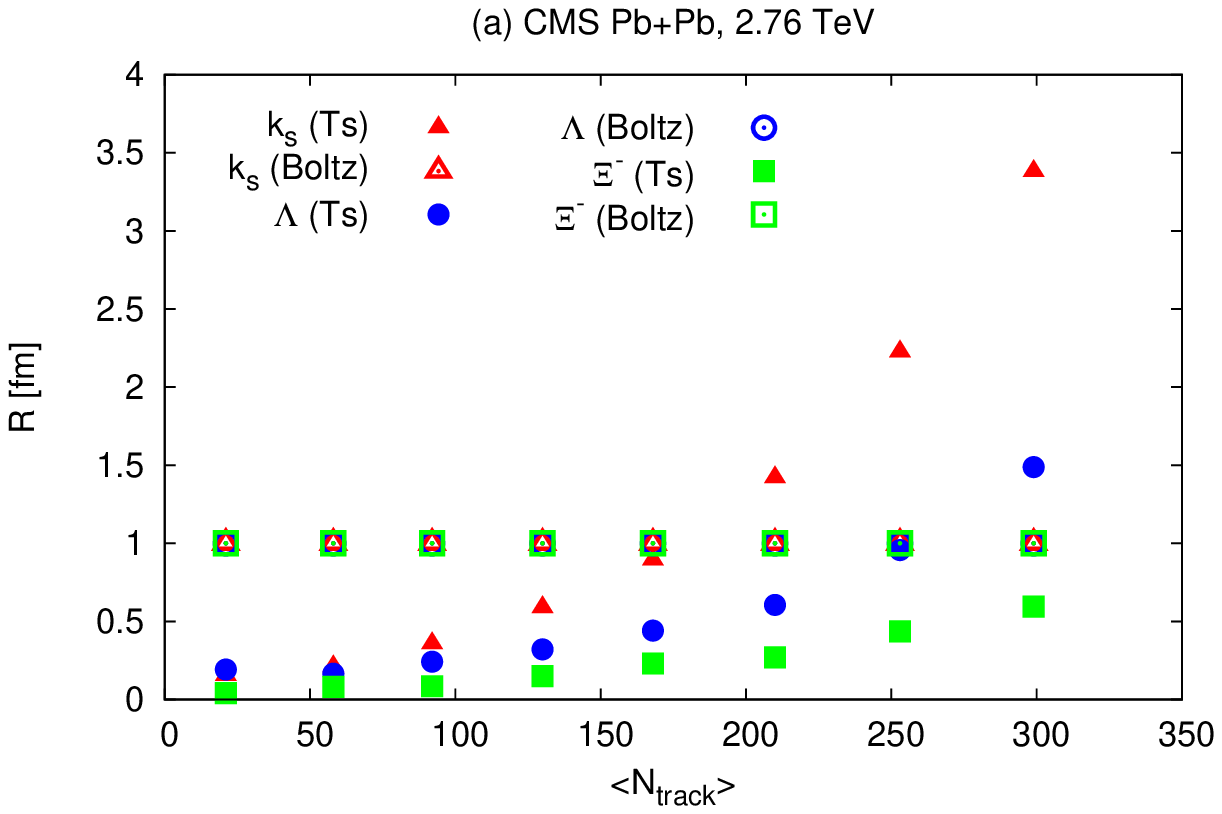}
\includegraphics[scale=0.65]{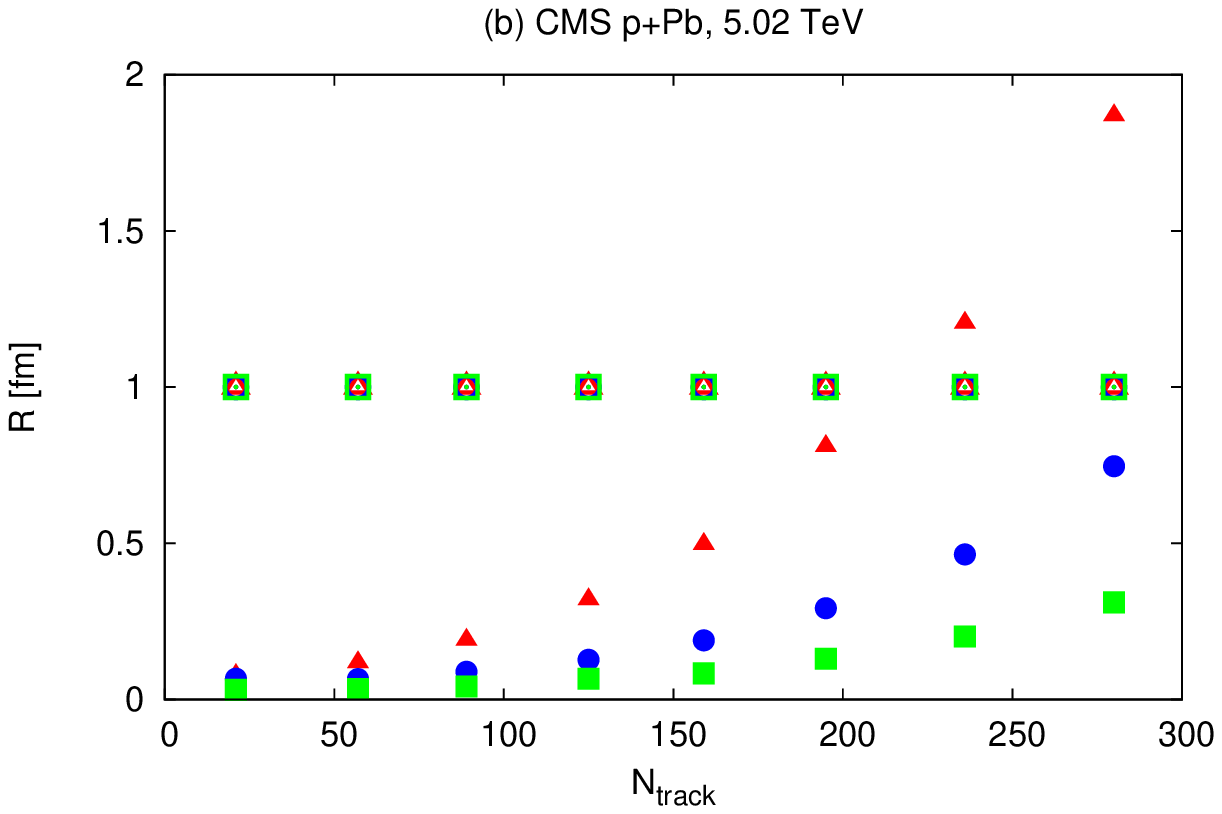}
\includegraphics[scale=0.65]{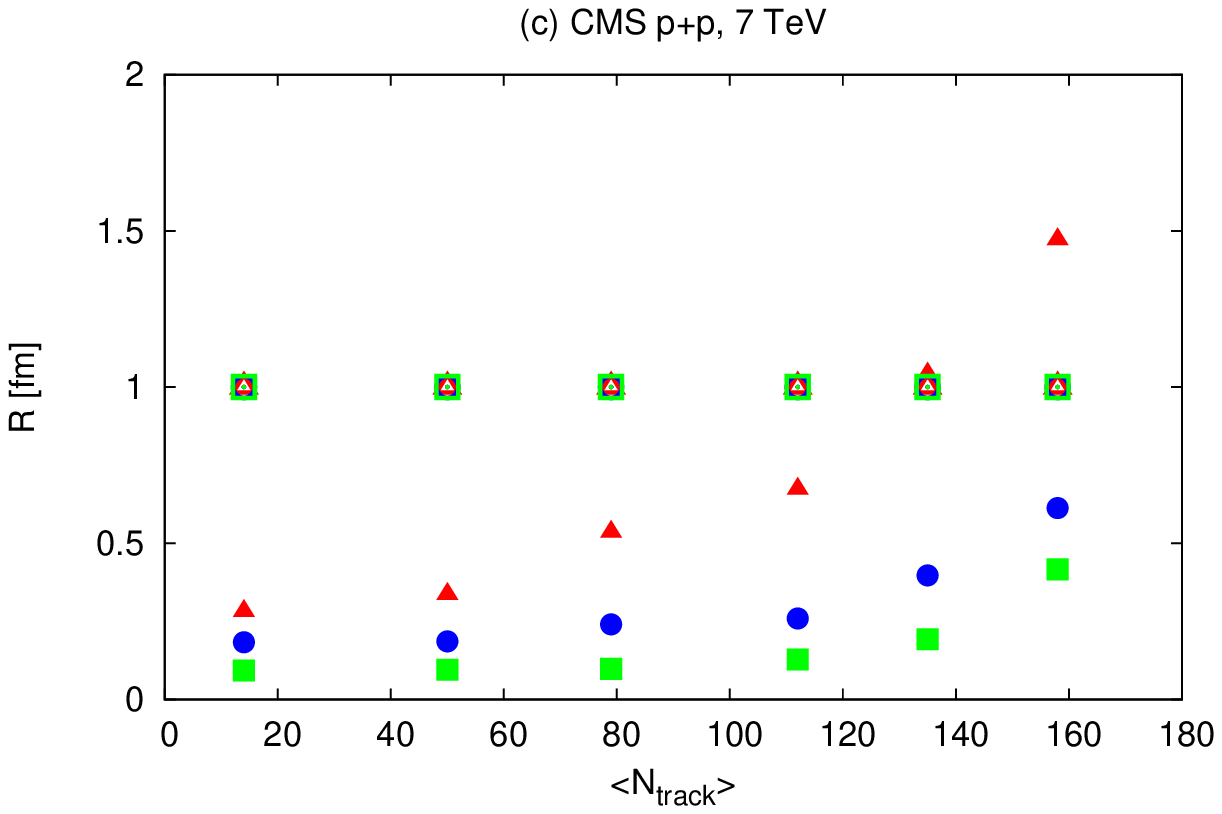}
\newline
\caption{(Color online) The fitting parameter $R$ as a function of the event multiplicity $<N_{\mathrm{track}}>$ for the strange particles \Kslxi which fitted using Tsallis and Boltzmann statistics for (a) $\textsf{Pb+Pb}$ collision at $\Sqsn=2.76$ TeV, (b) $\textsf{p+Pb}$ collision at $\Sqsn=5.02$ TeV, and (c) $\textsf{p+p}$ collision at $\Sqsn=7$ TeV.}
\label{comparision:R}
\end{center}
\end{figure*}

Figure~\ref{comparision:R} (a-c) show the fitting parameter $R$ for the strange particles \Kslxi as a function of $<N_{\mathrm{track}}>$ produced in the $\textsf{Pb+Pb}, \textsf{p+Pb},$ and $\textsf{p+p}$ collisions at $\sqrt{s_{\mathtt{NN}}}=2.76, 5.02,$ and $7$ TeV, respectively. The value of $R$ increases with the increasing in the multiplicity classes but decreases with the increase in the particle mass. So the volume of the system increases with the increase in the multiplicity events as expected.

\begin{figure*}[t!]
\begin{center}
\includegraphics[scale=0.65]{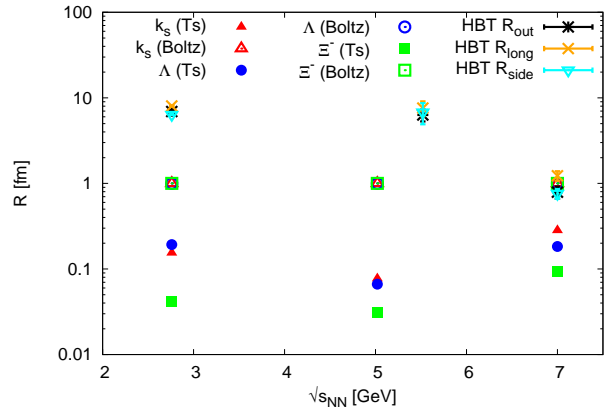}
\newline
\caption{(Color online) The parameter $R$ as a function of the center-of-mass energies $\Sqsn$ for the obtained strange particles \Kslxi which fitted using Tsallis and Boltzmann statistics and values of HBT radii ($R_{\mathrm{out}}, R_{\mathrm{long}}, R_{\mathrm{side}}$) \cite{Aamodt:2011mr,Okorokov:2014cna,Bialas:2014boa}.}
\label{R_HBT}
\end{center}
\end{figure*}

We compare our obtained values of $R$ with Hanbury-Brown–Twiss (HBT) radii \cite{Aamodt:2011mr,Okorokov:2014cna,Bialas:2014boa} at different center-of-mass energies $\sqrt{s_{\mathtt{NN}}}$ as shown in Fig.~\ref{R_HBT}. The fitting parameter $R$ for the strange particles \Kslxi and from HBT radii obtained at most central collisions and also at mid-rapidity (for details on HBT radius parameters, see Refs. \cite{Makhlin:1988,Boal:1990,Mayer:1992,Akkelin:1995,Chapman:1995}). The obtained values of $R$ were in the same behaviour as $R$ from HBT especially $R_{\mathrm{long}}$.

\begin{figure*}[t!]
\begin{center}
\includegraphics[scale=0.65]{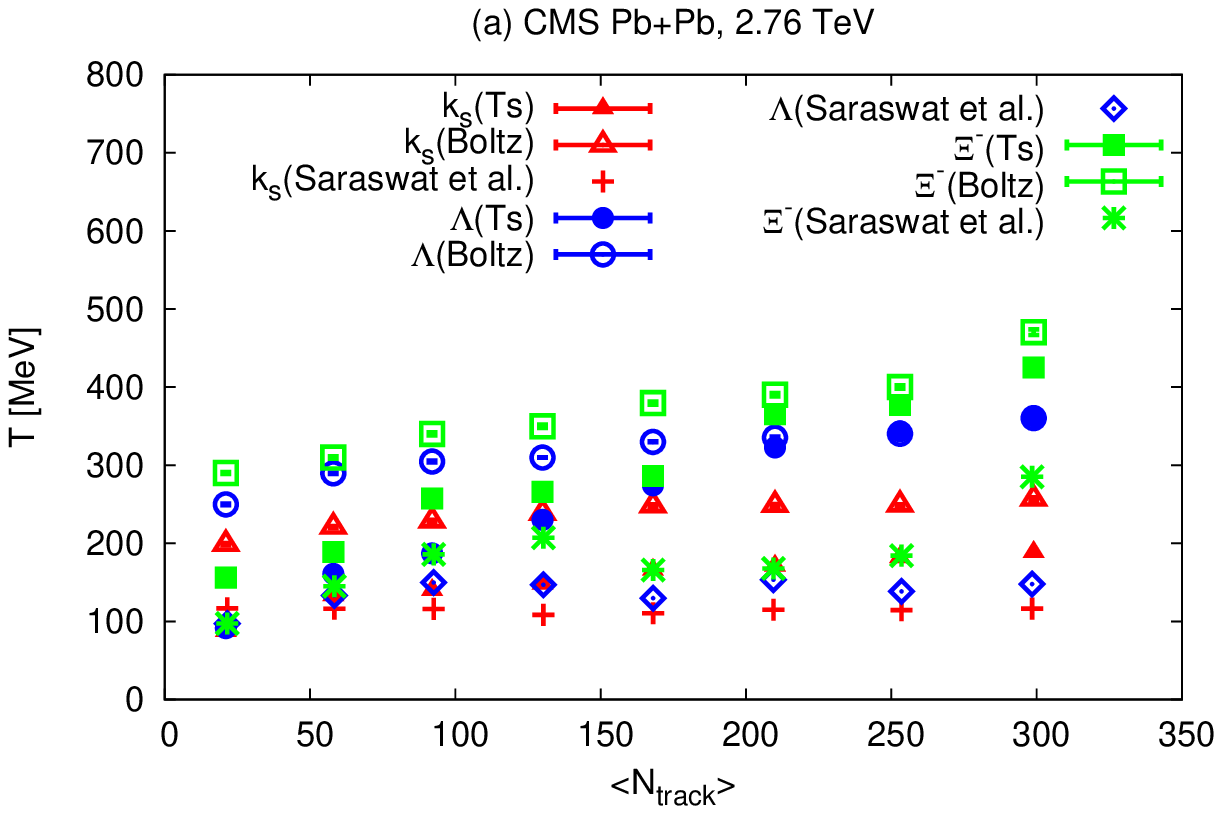}
\includegraphics[scale=0.65]{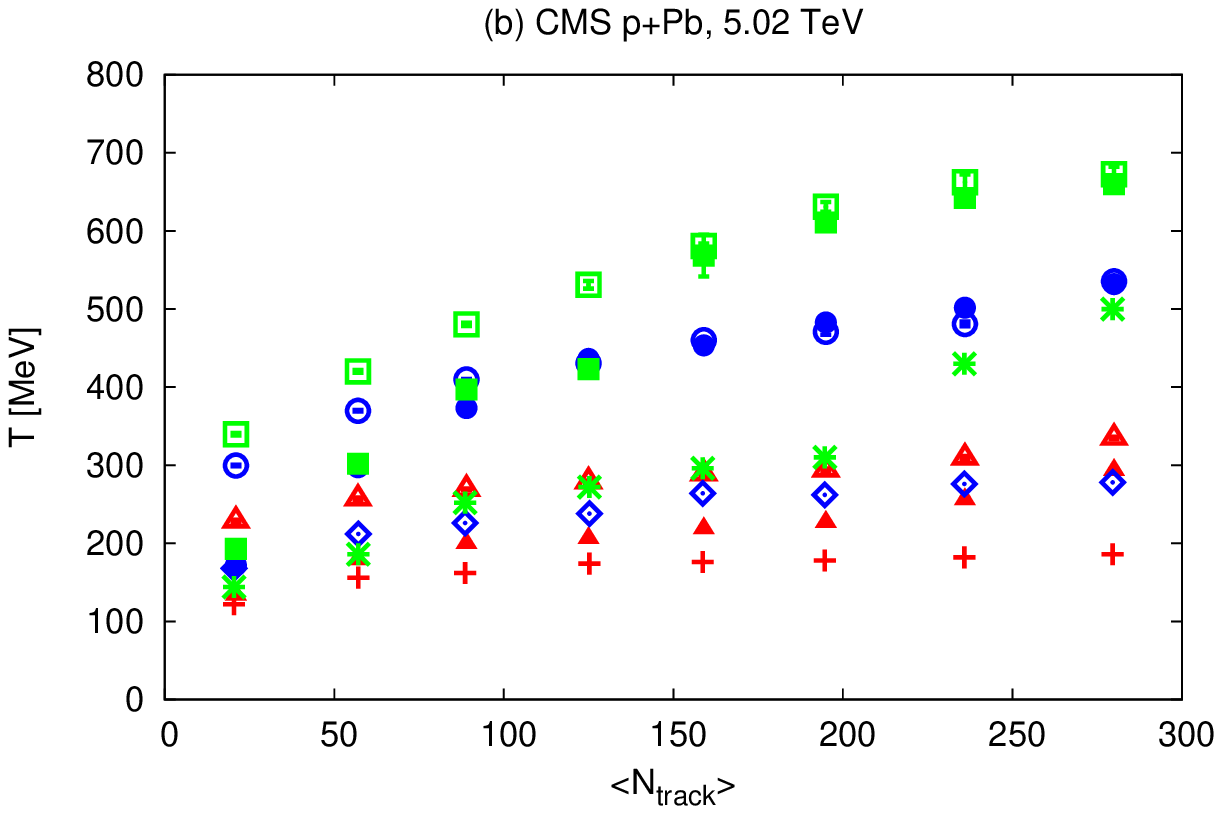}
\includegraphics[scale=0.65]{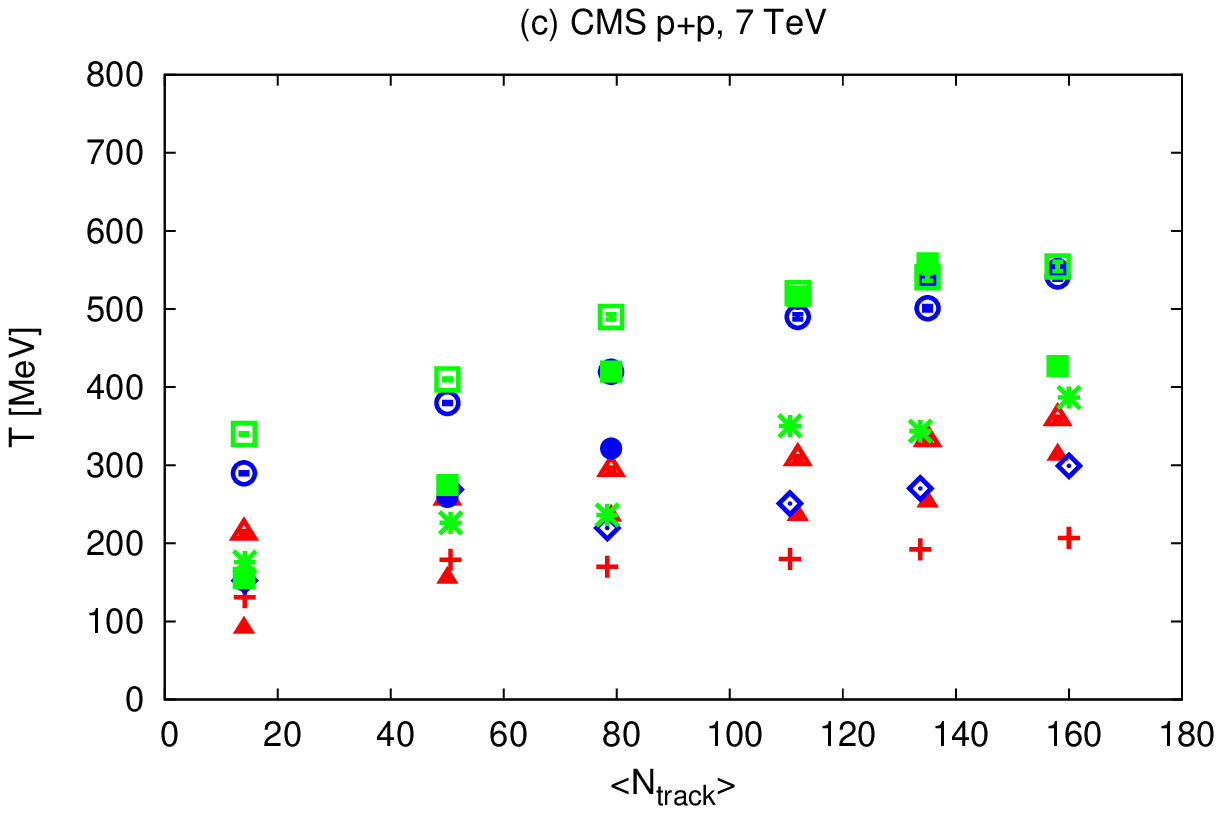}
\newline
\caption{(Color online) The fitting parameter $T$ as a function of the event multiplicity $<N_{\mathrm{track}}>$ for the strange particles \Kslxi which fitted using Tsallis and Boltzmann statistics- for (a) $\textsf{Pb+Pb}$ collision at $\Sqsn=2.76$ TeV, (b) $\textsf{p+Pb}$ collision at $\Sqsn=5.02$ TeV, and (c) $\textsf{p+p}$ collision at $\Sqsn=7$ TeV.}
\label{comparision:T}
\end{center}
\end{figure*}

Figure~\ref{comparision:T} (a-c) depict the fitting parameter $T$ for the strange particles \Kslxi as a function of $<N_{\mathrm{track}}>$ produced in the $\textsf{Pb+Pb}, \textsf{p+Pb},$ and $\textsf{p+p}$ collisions at $\sqrt{s_{\mathtt{NN}}}=2.76, 5.02,$ and $7$ TeV, respectively. The value of both temperatures (Tsallis and Boltzmann) increase with the increasing in both the particle mass and the strangeness number and also with the multiplicity classes for all systems. The direct relationship between the temperature and the strangeness number of hadrons is unlike the known behaviour obtained previously in extensive models (particle yields and ratios) \cite{Castorina:2014cia,Tawfik:2016tfe}, so we will work on the confirmation of the behaviour of this dependence in extensive and nonextensive particle yields and ratios in future work. Our calculations are compared with the modified Tsallis results \cite{Saraswat:2017gqt} and agreed with each other. The freeze-out of particles with large mass occur earlier than that of small mass. When volume is small, the particle with large mass freeze-out early \cite{Khuntia:2017ite}. Also, the Boltzmann temperature always greater than the Tsallis one which are independent on the type of particles or systems so $T_{\mathrm{Boltz}} > T_{\mathrm{Ts}}$.

\begin{figure}
\begin{center}
\includegraphics[scale=0.65]{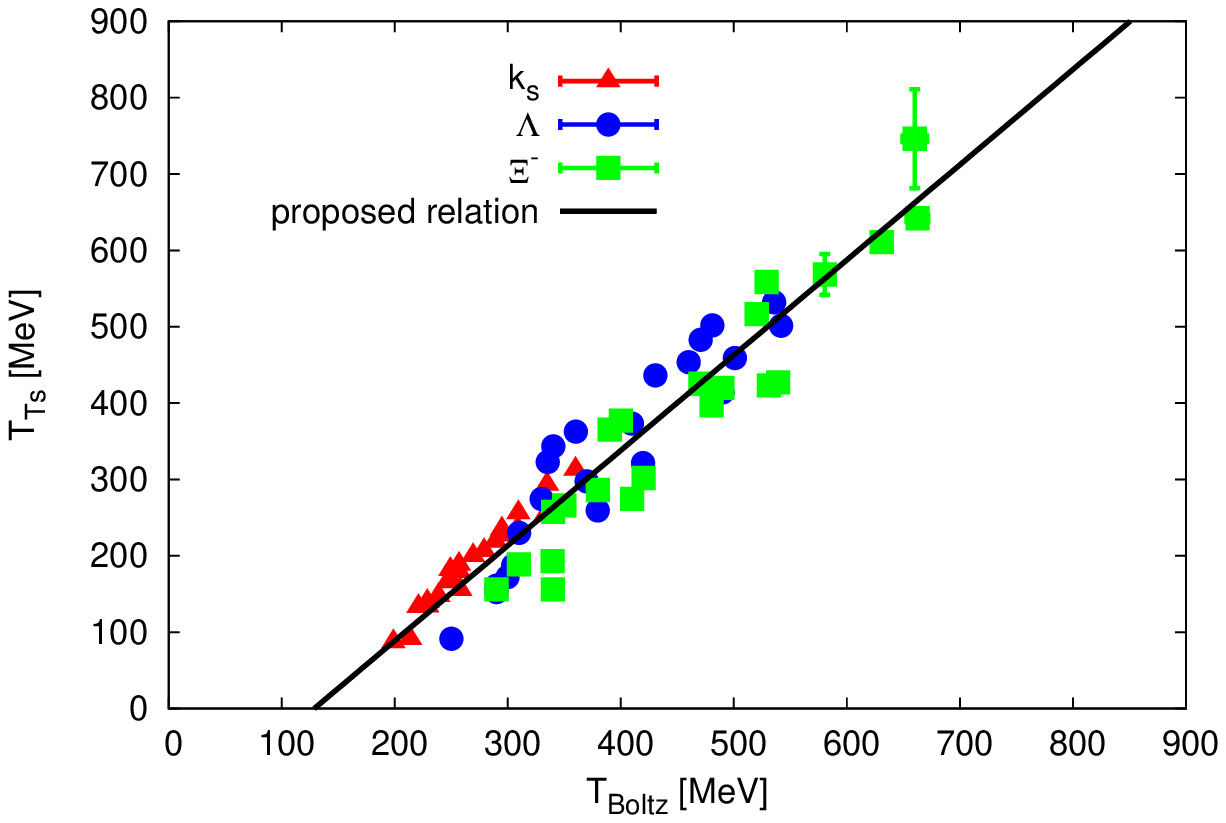}
\newline
\caption{(Color online) The Tsallis temperature $T_{\mathrm{Ts}}$ as a function of Boltzmann temperature $T_{\mathrm{Boltz}}$. All values of temperatures obtained from Figs. \ref{fit:Both:2.76}, \ref{fit:Both:5.02}, and \ref{fit:Both:7} which listed in Tables \ref{Tab1}, \ref{Tab2}, and \ref{Tab3} are given by symbols. The solid line represents the fitted result that given by Eq. \ref{TT:TB}.}
\label{TTsallis:TBoltzmann}
\end{center}
\end{figure}

%%%%%%%%%%%%%%%%%%%%%%%%%%%%%%%%%%%
\begin{landscape}
\begin{table*}[t]
\small
\begin{center}
\caption{Parameters deduced for the fitting of Tsallis and Boltzmann transverse momentum distributions at different multiplicity intervals for $\textsf{p+p}$ collision at $\sqrt{s_{\mathtt{NN}}}=7$ TeV, Fig. \ref{fit:Both:7}.}
\label{Tab3}
\begin{tabular}{ccccccccc}
%\toprule
\hline\noalign{\smallskip}
Particle & $<N_{\mathrm{track}}>$ & $T_{\mathtt{Ts}}$ [MeV] & $q$ & $R$ [fm] & $\chi^2/\mathrm{dof}$ & $T_{\mathtt{Boltz}}$ [MeV] & $R$ [fm] & $\chi^2/\mathrm{dof}$\\
\noalign{\smallskip}\hline\noalign{\smallskip}
$K_s^0$ & $14$ & $91.407 \pm 0.129$ & $1.154 \pm 0.001$ & $0.282 \pm 0.049$ & $0.026$ & $213.472 \pm 2.176$ & $0.126 \pm 0.04$ & $0.159$ \\
• & $50$ & $155.351 \pm 0.397$ & $1.1328 \pm 0.003$ & $0.337 \pm 0.066$ & $0.037$ & $258.135 \pm 2.336$ & $0.225 \pm 0.065$ & $0.123$ \\
• & $79$ & $234.716 \pm 1.78$ & $1.1047 \pm 0.013$ & $0.536 \pm 0.142$ & $0.093$ & $294.827 \pm 2.402$ & $0.463 \pm 0.125$ & $0.097$ \\
• & $111$ & $235.749 \pm 1.784$ & $1.1028 \pm 0.013$ & $0.673 \pm 0.178$ & $0.093$ & $308.689 \pm 2.798$ & $0.561 \pm 0.155$ & $0.105$ \\
• & $135$ & $252.973 \pm 1.445$ & $1.1082 \pm 0.011$ & $1.042 \pm 0.247$ & $0.067$ & $333.049 \pm 2.631$ & $0.869 \pm 0.226$ & $0.088$ \\
• & $158$ & $313.052 \pm 1.92$ & $1.0691 \pm 0.011$ & $1.472 \pm 0.355$ & $0.07$ & $359.903 \pm 2.047$ & $1.355 \pm 0.317$ & $0.064$ \\
\noalign{\smallskip}\hline\noalign{\smallskip}
  $\Lambda$ & $14$ & $152.631 \pm 0.284$ & $1.0945 \pm 0.0008$ & $0.183 \pm 0.039$ & $0.094$ & $289.682 \pm 1.218$ & $0.082 \pm 0.023$ & $0.209$ \\
• & $50$ & $259.374 \pm 0.211$ & $1.08 \pm 0.001$ & $0.186 \pm 0.028$ & $0.033$ & $379.636 \pm 1.303$ & $0.127 \pm 0.03$ & $0.136$ \\
• & $79$ & $321.4 \pm 1.805$ & $1.0714 \pm 0.005$ & $0.241 \pm 0.066$ & $0.202$ & $419.733 \pm 2.041$ & $0.193 \pm 0.05$ & $0.178$ \\
• & $111$ & $413.873 \pm 3.627$ & $1.0688 \pm 0.011$ & $0.309 \pm 0.093$ & $0.268$ & $490.01 \pm 3.286$ & $0.278 \pm 0.077$ & $0.216$ \\
• & $135$ & $459.049 \pm 2.757$ & $1.0519 \pm 0.008$ & $0.452 \pm 0.118$ & $0.176$ & $500.921 \pm 2.72$ & $0.434 \pm 0.113$ & $0.175$ \\
• & $158$ & $501.352 \pm 3.193$ & $1.0336 \pm 0.009$ & $0.661 \pm 0.173$ & $0.18$ & $541.774 \pm 4.08$ & $0.627 \pm 0.174$ & $0.213$ \\
\noalign{\smallskip}\hline\noalign{\smallskip}
  $\Xi^-$ & $14$ & $155.92 \pm 0.093$ & $1.099 \pm 0.0002$ & $0.093 \pm 0.014$ & $0.045$ & $339.888 \pm 0.995$ & $0.032 \pm 0.008$ & $0.203$ \\
• & $50$ & $274.655 \pm 0.151$ & $1.076 \pm 0.0003$ & $0.096 \pm 0.013$ & $0.034$ & $409.925 \pm 1.02$ & $0.062 \pm 0.014$ & $0.146$ \\
• & $79$ & $419.894 \pm 1.824$ & $1.0299 \pm 0.003$ & $0.098 \pm 0.026$ & $0.229$ & $489.854 \pm 3.369$ & $0.084 \pm 0.025$ & $0.345$ \\
• & $111$ & $516.464 \pm 1.844$ & $1.028 \pm 0.004$ & $0.128 \pm 0.03$ & $0.159$ & $520.321 \pm 2.446$ & $0.131 \pm 0.034$ & $0.225$ \\
• & $135$ & $528.961 \pm 2.024$ & $1.027 \pm 0.004$ & $0.204 \pm 0.048$ & $0.164$ & $540.85 \pm 4.672$ & $0.205 \pm 0.065$ & $0.403$ \\
• & $158$ & $539.162 \pm 3.965$ & $1.0204 \pm 0.008$ & $0.333 \pm 0.099$ & $0.326$ & $554.286 \pm 5.984$ & $0.329 \pm 0.111$ & $0.485$ \\
\noalign{\smallskip}\hline
%\bottomrule
\end{tabular}
\vspace{0mm}
\end{center}
\vspace{0mm}
\end{table*}
\end{landscape}
%%%%%%%%%%%%%%%%%%%%%%%%%%%%%%%%%%%
Fig.~\ref{TTsallis:TBoltzmann} shows the relation between $T_{\mathrm{Ts}}$ and $T_{\mathrm{Boltz}}$, where all values  of temperatures obtained from Figs. \ref{fit:Both:2.76}, \ref{fit:Both:5.02}, and \ref{fit:Both:7} which listed in Tables \ref{Tab1}, \ref{Tab2}, and \ref{Tab3} are given by symbols. The fitted result for all strange particles at different collisions is given by the solid line which is described by
\begin{equation}
T_{\mathrm{Ts}} = a \; \; T_{\mathrm{Boltz}} + b , \nonumber
\label{TT:TB}
\end{equation}
where $a=1.2465 \pm 0.0138$ and $b=-160.499 \pm 5.386$ are constants with $\chi^2/\mathrm{dof} = 0.0528$. But these values are different for different particles. So this linear relations is the same for all particles but the value of constants different depending on the mass of the particle. The dependence of the constants on the particle type is listed in Table \ref{Tab4}.

 %%%%%%%%%%%%%%%%%%%%%%%%%%%%%%%%%%%
\begin{table}
\begin{center}
\caption{Constants values deduced from the fitting of linear relation between Tsallis and Boltzmann temperatures, Fig. \ref{TTsallis:TBoltzmann}.}
\label{Tab4}
\begin{tabular}{cccc}
%\toprule
 \hline \noalign{\smallskip}
Particle & $a$ & $b$ & $\chi^2/\mathrm{dof}$ \\
\noalign{\smallskip}\hline%\noalign{\smallskip}
$K_s^0$ & $(1.3714 \pm 0.0092)$ & $(-177.514 \pm 2.503)$ & $0.0319$ \\
\noalign{\smallskip}\hline\noalign{\smallskip}
$\Lambda$ & $(1.3856 \pm 0.0255)$ & $(-209.84 \pm 10.2)$ & $0.0876$\\
\noalign{\smallskip}\hline\noalign{\smallskip}
$\Xi^-$ & $(1.39513 \pm 0.0212)$ & $(-249.213 \pm 9.963)$ & $0.0722$ \\
\noalign{\smallskip}\hline
%\bottomrule
\end{tabular}
\vspace{0mm}
\end{center}
\vspace*{0cm}
\end{table}
%%%%%%%%%%%%%%%%%%%%%%%%%%%%%%%%%%%
As can be seen from Table \ref{Tab4}, the value of constant $a$ increase with the increasing of the particle mass but constant $b$ inversely proportional to the particle mass. A similar linear relation between $T_{\mathrm{Ts}}$ and $T_{\mathrm{Boltz}}$ for charged (but not strange) particles at various collisions was found in Ref. \cite{Gao:2015qsq}. The values of constants $a$ and $b$ for the charged non-strange particles confirm our concept about the dependence of the constants on the particle mass.

Therefore, we can find a direct relation between any temperature and Boltzmann temperature and considering Boltzmann temperature as the $\textit{"reference"}$ for all other temperatures \cite{Gao:2015qsq}. However, Boltzmann statistics fails to fit some experimental transverse momentum spectra especially at high $p_{\mathrm{T}}$ region as a result of using the simplest form of Boltzmann statistics \cite{Schnedermann:1993ws,Meng:2009zzc,Zhao:2014fba,Gao:2015qsq}. On the contrary, Tsallis statistics succeeded in depicting all experimental transverse momentum spectra used in the present work and also at high region of $p_{\mathrm{T}}$ \cite{Saraswat:2017gqt,Gao:2015qsq,Cleymans:2012ya,35,36,37,38,39,40}. We also compare our Tsallis results with modified Tsallis one \cite{Saraswat:2017gqt} (with a radial flow). Our Tsallis temperatures are greater than modified one \cite{Saraswat:2017gqt}. The reason for this increasing gets from neglecting the effect of the radial flow. We neglect this parameter as a result of the considered low range of transverse momentum \cite{Gao:2015qsq}.
%%%%%%%%%%%%%%%%%%%%%%%%%%%%%%%%
%%%%%%%%%%%%%%%%%%%%%%%%%%%%%%%%
%%%%%%%%%%%%%%%%%%%%%%%%%%%%%%%%
%%%%%%%%%%%%%%%%%%%%%%%%%%%%%%%%
%%%%%%%%%%%%%%%%%%%%%%%%%%%%%%%%
%%%%%%%%%%%%%%%%%%%%%%%%%%%%%%%%
%%%%%%%%%%%%%%%%%%%%%%%%%%%%%%%%
%%%%%%%%%%%%%%%%%%%%%%%%%%%%%%%%
%%%%%%%%%%%%%%%%%%%%%%%%%%%%%%%%
%%%%%%%%%%%%%%%%%%%%%%%%%%%%%%%%

\section{Conclusion}
\label{sec:concl}
We analysed the transverse momentum $p_{\mathrm{T}}$ spectra of the strange particles \Kslxi in different multiplicity events produced in $\textsf{Pb+Pb}$ collision at $\Sqsn=2.76$ TeV, $\textsf{p+Pb}$ collision at $\Sqsn=5.02$ TeV, and $\textsf{p+p}$ collision at $\Sqsn=7$ TeV using Tsallis and Boltzmann statistics. In all cases nearly, our Tsallis and Boltzmann results are in excellent agreement with the experimental data of $\textit{CMS}$ collaboration at the LHC. Except for $K_s^0$, the goodness of our Boltzmann result decreases with the increase in the multiplicity events.

In all studied collisions, for all particles in different multiplicity events, the Tsallis $T_{\mathrm{Ts}}$ and Boltzmann $T_{\mathrm{Boltz}}$ temperatures values increase with the increasing in the mass of the mentioned particle. Also, there is a direct relationship between the values of the temperatures and the multiplicity events. We observe that $T_{\mathrm{Ts}} < T_{\mathrm{Boltz}}$, the value of $q$ is proportional inversely with the multiplicity events in all studied systems, and the value of $R$ in both statistics increases with the increase in the multiplicity events. As the direct dependence of the multiplicity on the size of the volume of the system.

There is a linear relation between the Tsallis temperature $T_{\mathrm{Ts}}$ and the Boltzmann temperature $T_{\mathrm{Boltz}}$. We have $T_{\mathrm{Ts}} = (1.3714 \pm 0.0092) \; \; T_{\mathrm{Boltz}} + (-177.514 \pm 2.503)$ for $K_s^0$, $T_{\mathrm{Ts}} = (1.3856 \pm 0.0255) \; \; T_{\mathrm{Boltz}} + (-209.84 \pm 10.2)$ for $\Lambda$, and  $T_{\mathrm{Ts}} = (1.39513 \pm 0.0212) \; \; T_{\mathrm{Boltz}} + (-249.213 \pm 9.963)$ for $\Xi^-$. Also, we have $T_{\mathrm{Ts}} = (1.2465 \pm 0.0138) \; \; T_{\mathrm{Boltz}} + (-160.499 \pm 5.386)$ for all studied strange particles. In all cases, $T_{\mathrm{Ts}} < T_{\mathrm{Boltz}}$ as the Boltzmann temperature is considered as the base for all other temperatures, and both values of temperatures increase with the increasing in the particle mass at vanishing chemical potential.
%\appendix

%\section{Appendices}

%\clearpage


\begin{thebibliography}{99}
\bibitem{Bjorken:1982qr} J.D. Bjorken, %{Highly Relativistic Nucleus-Nucleus Collisions: The Central Rapidity Region}",
Phys. Rev. D {\bf 27}, 140 (1983).
 %     doi            = "10.1103/PhysRevD.27.140"}

\bibitem{Ullrich:2013qwa} T. Ullrich, B. Wyslouch, J.W. Harris, %Proceedings, 23rd International Conference on Ultrarelativistic Nucleus-Nucleus Collisions : Quark Matter 2012 (QM 2012)}",
Nucl. Phys. A {\bf 904-905}, 1c (2013).

\bibitem{Gyulassy:2004zy} M. Gyulassy, L. McLerran, %{New forms of QCD matter discovered at RHIC}",
Nucl. Phys. A {\bf 750}, 30 (2005).
%      doi            = "10.1016/j.nuclphysa.2004.10.034", "nucl-th/0405013",

\bibitem{Khuntia:2017ite} A. Khuntia, S. Tripathy, R. Sahoo, J. Cleymans,
%{Multiplicity Dependence of Non-extensive Parameters for Strange and Multi-Strange Particles in Proton-Proton Collisions at $\sqrt{s}= 7$ TeV at the LHC}",
Eur. Phys. J. A {\bf 53}, 103 (2017).
%doi            = "10.1140/epja/i2017-12291-8","1702.06885", "hep-ph"

\bibitem{Ahmad:2013una} S. Ahmad, A. Ahmad, A. Chandra, M. Zafar, M. Irfan,
%title          = "{Entropy Analysis in Relativistic Heavy-Ion Collisions}",
Adv. High Energy Phys. {\bf 2013}, 836071 (2013).
%      doi            = "10.1155/2013/836071",

\bibitem{Tawfik:2014eba} A. Tawfik,
%”Equilibrium statistical-thermal models in high-energy”,
Int. J. Mod. Phys. A {\bf 29}, 1430021 (2014).

\bibitem{Rafelski:1994gi}  J. Letessier, J. Rafelski, A. Tounsi, Phys. Rev. C {\bf 50}, 406 (1994).

\bibitem{ALICE:2017jyt} J. Adam et al. (ALICE collaboration),
%{Enhanced production of multi-strange hadrons in high-multiplicity proton-proton collisions}",
Nature Phys. {\bf 13}, 535 (2017).
%doi            = "10.1038/nphys4111", "1606.07424", nucl-ex",

\bibitem{Rafelski:1982pu} J. Rafelski, B. Muller, %"{Strangeness Production in the Quark - Gluon Plasma}",
Phys. Rev. Lett. {\bf 48}, 1066 (1982).
%doi            = "10.1103/PhysRevLett.48.1066",
Erratum: Phys. Rev. Lett. {\bf 56}, 2334 (1986).

\bibitem{Song:2010mg} H-C Song, S.A. Bass, U.W. Heinz, T. Hirano, C. Shen,
Phys. Rev. Lett. {\bf 106}, 192301 (2011), Erratum: Phys. Rev. Lett. {\bf 109}, 139904 (2012).

\bibitem{Adams:2003xp} J. Adams et al. (STAR Collaboration), Phys. Rev. Lett. {\bf 92}, 112301
(2004).

\bibitem{Becattini:2004td} F. Becattini, J. Phys. Conf. Ser. {\bf 5}, 175 (2005).

\bibitem{Mueller:1993rr} A.H. Mueller, Nucl. Phys. B {\bf 415}, 373 (1994).

\bibitem{Mueller:1994jq} A.H. Mueller, B. Patel, Nucl. Phys. B {\bf 425}, 471 (1994).

\bibitem{Mueller:1994gb} A.H. Mueller, Nucl. Phys. B {\bf 437}, 107 (1995).

\bibitem{Schnedermann:1993ws} E. Schnedermann, J. Sollfrank, U.W. Heinz, Phys. Rev. C {\bf 48}, 2462 (1993).

\bibitem{Tawfik:2017bul} A.N. Tawfik, H. Yassin, E.R. Abo Elyazeed, Chin. Phys. C {\bf 41}, 053107 (2017).
 %     doi            = "10.1088/1674-1137/41/5/053107",  = "1701.04697"

\bibitem{Tsallis:1987eu} C. Tsallis, J. Statist. Phys. {\bf 52}, 479 (1988).

\bibitem{Biro:2008hz} T.S. Biro, G. Purcsel, K. Urmossy, Eur. Phys. J. A {\bf 40}, 325  (2009). %[arXiv:0812.2104 [hep-ph]].


\bibitem{22} T. Bhattacharyya, J. Cleymans, A. Khuntia, P. Pareek, R. Sahoo, Eur. Phys. J. A {\bf 52}, 30 (2016).

\bibitem{23} H. Zheng, L. Zhu, Adv. High Energy Phys. {\bf 2015}, 180491 (2015).

\bibitem{24} Z. Tang, Y. Xu, L. Ruan, G. van Buren, F. Wang, Z. Xu, Phys. Rev. C {\bf 79}, 051901 (2009).

\bibitem{25} B. De, Eur. Phys. J. A {\bf 50}, 138 (2014).

\bibitem{Adare:2011vy} A. Adare et al. (PHENIX Collaboration), Phys. Rev. C {\bf 83}, 064903 (2011).

\bibitem{Khachatryan:2011tm} V. Khachatryan et al. (CMS Collaboration), JHEP {\bf 1105}, 064 (2011).

\bibitem{Khandai:2013gva} P.K. Khandai, P. Sett, P. Shukla, V. Singh, Int. J. Mod. Phys. A {\bf 28}, 1350066 (2013). %[arXiv:1304.6224 [hep-ph]].

\bibitem{Adare:2010fe} A. Adare et al. (PHENIX Collaboration), Phys. Rev. D {\bf 83}, 052004 (2011).
%[arXiv:1005.3674 [hep-ex]].

\bibitem{Sett:2014csa} P. Sett, P. Shukla, Adv. High Energy Phys. {\bf 2014}, 896037 (2014).
% [arXiv:1408.1034 [hep-ph]].

\bibitem{Khandai:2013fwa} P.K. Khandai, P. Sett, P. Shukla, V. Singh, J. Phys. G {\bf 41}, 025105 (2014).
% [arXiv:1310.4022 [nucl-th]].

\bibitem{Saraswat:2017gqt} K. Saraswat, P. Shukla, V. Kumar, V. Singh, %"{Strange hadron production in pp, pPb and PbPb collisions at LHC energies}",
Eur. Phys. J. A {\bf 53}, 5 (2017).
%doi    = "10.1140/epja/i2017-12276-7","1702.05734","nucl-th"

\bibitem{21} J. Cleymans, D. Worku, J. Phys. G {\bf 39}, 025006 (2012).

\bibitem{Sett:2015lja} P. Sett, P. Shukla,
%"{Inferring freeze-out parameters from pion measurements at RHIC and LHC}",
Int. J. Mod. Phys. E {\bf 24}, 1550046 (2015).
%doi   = "10.1142/S0218301315500469","1505.05258", hep-ph

\bibitem{Gao:2015qsq} Y-Q Gao, F-H Liu, %"{Comparing Tsallis and Boltzmann temperatures from relativistic heavy ion collider and large hadron collider heavy-ion data}",
Indian J. Phys. {\bf 90}, 319 (2016).
%doi   = "10.1007/s12648-015-0747-z",

\bibitem{Tsallis:1998ws} C. Tsallis, R.S. Mendes, A.R. Plastino, Physica A {\bf 261}, 534 (1998).

\bibitem{Cleymans:2012ya} J. Cleymans, D. Worku, Eur. Phys. J. A {\bf 48}, 160 (2012).

\bibitem{46} J. Cleymans, M.D. Azmi, Eur. Phys. J. C {\bf 75}, 430 (2015).

\bibitem{Cleymans:2015lxa} J. Cleymans, M.D. Azmi, %{Large Transverse Momenta and Tsallis Thermodynamics}",
J. Phys. Conf. Ser. {\bf 668}, 012050 (2016).
%doi   = "10.1088/1742-6596/668/1/012050", "1508.03143","hep-ph",

\bibitem{47} B.C. Li, Z. Zhang, J.H. Kang, G.X. Zhang, F.H. Liu, Adv. High Energy Phys. {\bf 2015}, 741816 (2015).


\bibitem{2} P-Z Ning, L. Li, D-F Min, \textit{Foundation of Nuclear Physics: Nucleons and Nuclei}, % $2^{nd}$ edn.
(Higher Education Press, Beijing, China, 2003).

\bibitem{3} C.D. Dermer, Astrophys. J. {\bf 280}, 328 (1984).

\bibitem{Meng:2009zzc} C-R Meng, Chin. Phys. Lett. {\bf 26}, 102501 (2009).

\bibitem{Khachatryan:2016yru} V. Khachatryan et al. (CMS Collaboration),
%"{Multiplicity and rapidity dependence of strange hadron production in pp, pPb, and PbPb collisions at the LHC}",
Phys. Lett. B {\bf 768}, 103 (2017). %[arXiv:1605.06699 [nucl-ex]].

\bibitem{Chatrchyan:2013nka} S. Chatrchyan et al. (CMS Collaboration),
%"{Multiplicity and transverse momentum dependence of two-and four-particle correlations in pPb and PbPb  collisions}",
Phys. Lett. B {\bf 724}, 213 (2013).
% doi   = "10.1016/j.physletb.2013.06.028","1305.0609", "nucl-ex",


\bibitem{50} J. Cleymans, G.I. Lykasov, A.S. Parvan, A.S. Sorin, O.V. Teryaev, D. Worku, Phys. Lett. B {\bf 723}, 351 (2013).

\bibitem{Aamodt:2011mr} K. Aamodt et al. (ALICE Collaboration),
  %``Two-pion Bose-Einstein correlations in central Pb-Pb collisions at $\sqrt{{s}_{NN}} =$ 2.76 TeV,''
  Phys.\ Lett.\ B {\bf 696}, 328 (2011).
%doi:10.1016/j.physletb.2010.12.053 [arXiv:1012.4035[nucl-ex]].

\bibitem{Okorokov:2014cna} V.A. Okorokov,
  %``Azimuthally integrated HBT parameters for charged pions in nucleus-nucleus interactions versus collision energy,''
Adv.\ High Energy Phys.\  {\bf 2015}, 790646 (2015).
%doi:10.1155/2015/790646 [arXiv:1409.3925 [nucl-ex]].

\bibitem{Bialas:2014boa} A. Bialas, W. Florkowski, K.Zalewski,
  %``Blast-wave model description of the Hanbury-Brown-Twiss radii in $pp$ collisions at LHC energies,''
J.\ Phys.\ G {\bf 42}, no. 4, 045001 (2015).
%doi:10.1088/0954-3899/42/4/045001 [arXiv:1409.5240 [hep-ph]]

\bibitem{Makhlin:1988} A.N. Makhlin, Yu.M. Sinyukov, Z. Phys. C {\bf 39}, 69 (1988).

\bibitem{Boal:1990} D. Boal, C.G. Gelbke, B. Jennings, Rev. Mod. Phys. {\bf 62}, 553 (1990).

\bibitem{Mayer:1992} U. Mayer, E. Schnedermann, U. Heinz, Phys. Lett. B {\bf 294}, 69 (1992).

\bibitem{Akkelin:1995} S.V. Akkelin, Yu.M. Sinyukov, Phys. Lett. B {\bf 356}, 525  (1995).

\bibitem{Chapman:1995} S. Chapman, P. Scotto, U. Heinz, Phys. Rev. Lett. {\bf 74}, 440 (1995).

\bibitem{Castorina:2014cia} P. Castorina, H. Satz,
  %``Hawking-Unruh Hadronization and Strangeness Production in High Energy Collisions,''
Adv.\ High Energy Phys.\  {\bf 2014}, 376982 (2014).
%doi:10.1155/2014/376982  [arXiv:1403.3541 [hep-ph]].

\bibitem{Tawfik:2016tfe} A.N. Tawfik, H. Yassin, E.R.A. Elyazeed,
  %``Strangeness production in high-energy collisions and Hawking-Unruh radiation,''
  Int.\ J.\ Mod.\ Phys.\ E {\bf 26}, no. 03, 1750001 (2017).
%doi:10.1142/S021830131750001X   [arXiv:1612.05105 [physics.gen-ph]].


\bibitem{Zhao:2014fba} H. Zhao, F-H Liu, %"{Chemical Potentials of Quarks Extracted from Particle Transverse Momentum Distributions in Heavy Ion Collisions at RHIC Energies}",
Adv. High Energy Phys. {\bf 2014}, 742193 (2014).
%      doi            = "10.1155/2014/742193",


\bibitem{35} C-Y Wong, G. Wilk, Acta. Phys. Pol. B {\bf 43}, 2047 (2012).

\bibitem{36} K. Urmosy, G.G. Barnafoldi, T.S. Biro, Phys. Lett. B {\bf 718}, 125 (2012).

\bibitem{37} G. Wilk, Z. Wldarczyk, Eur. Phys. J. A {\bf 48}, 161 (2012).

\bibitem{38} L. Marques, J. Cleymans, A. Deppman, Phys. Rev. D {\bf 91}, 054025 (2015).

\bibitem{39} L. Marques, E. Andrade-II, A. Deppman, Phys. Rev. D {\bf 87}, 114022 (2013).

\bibitem{40} M.D. Azmi, J. Cleymans, J. Phys. G {\bf 41}, 065001 (2014).

%-------------------------------------------------------dfo---------------------------------------------------------------------------

 \end{thebibliography}
\end{document}